\documentclass[10pt, journal]{IEEEtran}

\usepackage{epsfig}
\usepackage{amsmath}
\usepackage{amssymb}
\usepackage{amsfonts}
\usepackage{subfigure}
\usepackage{mathtools}
\usepackage{balance}
\usepackage{lipsum}
\usepackage{tabularx}
\usepackage{graphicx}
\usepackage[outdir=./]{epstopdf}
\usepackage{stfloats}
\usepackage[utf8x]{inputenc}
\usepackage{bm}
\usepackage{color}
\usepackage{cite}
\usepackage{textcomp}
\usepackage{bigints}
\usepackage{hyperref}
\usepackage[many]{tcolorbox}

\newcommand{\vc}[3]{\overset{#2}{\underset{#3}{#1}}}
\newcommand*{\QED}{\hfill\ensuremath{\square}}
\newcommand*{\QEDB}{\hfill\ensuremath{\blacksquare}}%

\DeclareMathOperator*{\argmax}{arg\,max}
\DeclareMathOperator*{\bfB}{\textbf{B}}
\DeclareMathOperator*{\bfP}{\textbf{P}}
\DeclareMathOperator*{\bfp}{\textbf{p}}
\DeclareMathOperator*{\bfb}{\bm{\beta}}

\newtheorem{theorem}{Theorem}
\newtheorem{lemma}{Lemma}

\newtheorem{definition}{Definition}

\AtBeginEnvironment{subappendices}{%
\chapter*{Appendix}
\addcontentsline{toc}{chapter}{Appendices}
\counterwithin{figure}{section}
\counterwithin{table}{section}
}

\allowdisplaybreaks

\begin{document}

\title{Double-Auction Mechanisms for \\ Resource Trading Markets}

\author{K.~P.~Naveen,~\IEEEmembership{Member,~IEEE}  and Rajesh Sundaresan,~\IEEEmembership{Senior Member,~IEEE} 
\thanks{K.~P.~Naveen is with the Department of Electrical Engineering, Indian Institute of Technology Tirupati, India (email: naveenkp@iittp.ac.in).}
\thanks{R.~Sundaresan is with the Department of Electrical Communication Engineering, Robert Bosch Centre for Cyber Physical Systems and the Centre for Networked Intelligence, at the Indian Institute of Science, Bangalore, India (email: rajeshs@iisc.ernet.in).}
\thanks{The work of the first author was supported by an INSPIRE Faculty Award (No. DST/INSPIRE/04/2015/000928) of the Department of Science and Technology, and a Startup Research Grant (No. SRG/2019/001693) of the Science Education and Research Board, both of the Government of India. The second author was supported by the Robert Bosch Centre for Cyber-Physical Systems and the Centre for Networked Intelligence, Indian Institute of Science, Bangalore. This paper was presented in part at the 16th International Symposium on Modeling and Optimization in Mobile, Ad Hoc, and Wireless Networks (WiOpt 2018) \cite{naveen2018double}.}
}

\maketitle

\begin{abstract}
We consider a double-auction mechanism, which was recently proposed in the context of rate allocation in mobile data-offloading markets; our mechanism is also applicable to the problem of bandwidth allocation in network slicing markets. Network operators (\emph{users}) derive benefit from offloading their traffic to third party WiFi or femtocell networks (\emph{link-suppliers}). Link-suppliers experience costs for the additional capacity that they provide. Users and link-suppliers (collectively referred to as \emph{agents}) have their pay-offs and cost functions as private knowledge. A \emph{network-manager}  decomposes the problem into a network problem (with surrogate pay-offs and surrogate cost functions) and agent problems (one per agent). The surrogate pay-offs and cost functions are modulated by the agents' bids. Agents' payoffs and costs are then determined by the allocations and prices set by the network-manager. Under this design, so long as the agents do not anticipate the effect of their actions on the prices set by the network-manager (i.e., \emph{price-taking} agents), a competitive equilibrium exists as a solution to the network and agent problems, and this equilibrium optimizes the sum utility of all agents. However, this design fails when the agents (including the link-supplier) are all strategic (\emph{price-anticipating}). Specifically, the presence of a strategic link-supplier drives the system to an undesirable equilibrium with zero participation resulting in an efficiency loss of 100\%. This is in stark contrast to an earlier setting where the users alone are strategic but the link-supplier is not $-$ the efficiency loss is known to be at most 34\%. The paper then proposes the following  Stackelberg game modification with asymmetric information structures for link-supplier and users in order to alleviate the efficiency-loss problem: the network-manager  first announces the allocation and payment functions; he then invites the link-supplier to announce its bid, following which the  users are invited to respond with their bids. The resulting Stackelberg games' efficiency losses can be characterized in terms of the link-supplier's  cost function when the users' pay-off functions are linear. Specifically, when the link-supplier's cost function is quadratic, the worst case efficiency loss is 25\%. Further, the loss in efficiency improves for polynomial cost functions of higher degree. For non-linear utility functions (e.g., $\alpha$-fair and $\log$ utilities), we demonstrate the efficacy of the  proposed mechanism via. a detailed numerical study. 
\end{abstract}

\begin{IEEEkeywords}
Network utility maximization, double-auction, KKT conditions, Nash equilibrium, Stackelberg equilibrium.
\end{IEEEkeywords}

\section{Introduction}
We consider double auction mechanisms motivated by two recent proposals -- {mobile data offloading} and network slicing-based virtualization. Mobile data offloading is an effective way to manage growth in mobile-data traffic. Traffic meant for the macrocellular network can be offloaded to already installed third-party Wi-Fi or femtocell networks. This provides an alternative means of network expansion. Wi-Fi access-point operators and femtocell network operators will however expect compensation for allowing macrocellular network traffic through their access points. Technological, security, and preliminary economic studies for secure and seamless offloading have been discussed in \cite{european_study,cisco,duan_etal13economics,duan_etal16capacity_allocation}. Network slicing \cite{faukas_etal17network_slicing} on the other hand is a virtualization technique that allows many logical networks to run atop shared physical networks. It allows physical mobile network operators to partition their network resources and offer them to different users or tenants (IoT streams, mobile broadband streams, etc.) in return for suitable compensation. It enables network operators to focus on their core strength of delivering high-quality network experiences while the tenants or virtual network operators can focus more on business, billing, and branding relations.

Although the above two proposals are technologically different, the underlaying resource allocation problems are similar. Both examples comprise buyers or \emph{users} (mobile operators in data-offloading scenario and logical-network operators in network-slicing)  who intend to purchase a portion of the resource (rate or bandwidth) available with a \emph{supplier} (femtocell and physical-network operator, respectively).  In this work, using the framework of \emph{Network Utility Maximization (NUM)}, we propose \emph{double-auction mechanisms} for trading resources in such markets. In the remainder of the paper, for the ease of exposition, we discuss our work in the context of mobile data offloading;  the mapping to the context of network slicing will be obvious. We proceed by discussing related literature and then present our main technical contributions. 

\subsection{Related Work}
The traditional model of  network utility maximization (NUM)   proposed by Kelly in his seminal work \cite{kelly97charging-rate-control} comprises a collection of users and a network-manager (but not a supplier like in our case). The network-manager's objective is to allocate rates to users (over a link\footnote{In general, it can be a network of links; we adhere to the single link case so that we can draw a comparison with our model.}) in a distributed fashion so as to maximize the sum-utility of the users. Over the years several interesting variants of the traditional model has been proposed. For instance, the problem of distributed implementation was studied by Kelly et al.  \cite{kelly-etal98rate-control}, while Yang and Hajek \cite{yang_hajek07VCG_kelly_mechanism} and Johari and Tsitsiklis in \cite{johari_tsitsiklis05VCG_mechanism,johari-tsitsiklis04efficiency-loss} consider the problem of characterizing the loss in efficiency that is incurred when the users are strategic. For further literature on the topic, see \cite{yi_chiang08stochastic_NUM} and references therein.

In the traditional setup, although the allocated rate is supplied to the users over a link, the link itself is however  considered to be a passive element. This is in contrast to our model where the link is assumed to be maintained by a strategic link-supplier. In the context of mobile data offloading, the femtocell operator can be regarded as the link-supplier while the cellular operators (seeking to offload some of their subscribers onto the femtocell) constitute the users. Similarly, in network slicing the physical-network operator (owning the bandwidth) is the link-supplier, while the  logical-network operators (who intend to rent a portion of the link-supplier's bandwidth) are the users. A neutral network-manager such as a regulatory authority facilitates the exchange of resources (rates or bandwidth) between the link-supplier and the users. 

For the above setup, Iosifidis et al. \cite{iosifidis-etal15double-auction} in their recent work propose a double auction mechanism which achieves the system  objective of sum-utility maximization. Specifically, the mechanism works as follows. The network-manager collects information about how much each user is willing to pay for the link-supplier's resource. Simultaneously, the link-supplier provides bids (to the network-manager) that contains information about the amount of resource that the link-supplier is willing to share with the individual users. Using these scalar signals, the network-manager then allocates resources (rates) to the users, along with determining the payments that each user has to make to the link-supplier. The users and the link-supplier then comply. Following Kelly et al.\ \cite{kelly-etal98rate-control}, Iosifidis et al. \cite{iosifidis-etal15double-auction} showed that a t\^{a}tonnement procedure converges to the system optimal operating point.

The work in \cite{iosifidis-etal15double-auction} is however limited to scenarios where the agents are only \emph{price-taking}\footnote{Price-taking agents simply accept the prices communicated by the network-manager without investigating  the process via. which the prices may have been set (details are available in Section~\ref{price_taking_section}).}. In the current work, we extend their double-auction mechanism to incorporate  \emph{price-anticipating agents}\footnote{Price-anticipating agents take actions by anticipating the effect of their actions on the prices set by the network-manager (see Section~\ref{price_anticipating_section} for details).}. Specifically, we propose a Stackelberg-game based double-auction mechanism, and prove a lower bound on the loss in efficiency (incurred because of the price-anticipating nature of the agents). Finally, from the context of double-auctions, Iosifidis et al.\ \cite[p.1635]{iosifidis-etal15double-auction} point out that designing incentive compatible mechanisms for double-auctions which are weakly budget balanced (i.e., the broker should not end up subsidizing the mechanism) is `notoriously hard' and has been done only in certain simplified settings (McAfee auction \cite{mcafee92dominant_strategy}) or can be computationally intensive.  So \cite{iosifidis-etal15double-auction} took a network utility maximization approach and left the analysis of the price-anticipating scenario open \cite[Sec~VII, p.1646]{iosifidis-etal15double-auction}. In this work we address the above gap in literature.

\subsection{Our Contributions}
Our contributions in this paper are as follows.
\begin{enumerate}
\item We first re-derive the result on efficient allocation when the agents are price-taking \cite{iosifidis-etal15double-auction}, mainly to set up the notation for the subsequent results (Section~\ref{price_taking_section}). The solution is characterized in terms of a \emph{competitive equilibrium} whose optimality (in terms of maximizing the sum-utility) is already known from \cite{iosifidis-etal15double-auction}.  However, the proof of optimality in \cite{iosifidis-etal15double-auction} is based on learning dynamics; we instead provide an alternate proof that is based on Lagrangian techniques.

\item We then analyze the price-anticipating scenario along the lines of Johari et al.\ \cite{johari-etal2005efficiency} in Section~\ref{price_anticipating_section}. The solution is characterized in terms of a \emph{Nash equilibrium}. The situation in Johari et al.\ \cite{johari-etal2005efficiency}, when mapped to the current offloading setting, would be one where the users alone are strategic, while the link is simply a passive entity; the efficiency loss due to price-anticipating users is then known to be at most 34\% \cite{johari-etal2005efficiency}. However, as in our case, when the link-supplier is also strategic and price-anticipating, we find that the efficiency loss is aggravated to 100\%. Thus, the double-auction mechanism of \cite{iosifidis-etal15double-auction} (designed for price-taking agents) fails in the more practical scenario where the agents are price-anticipating.

\item Motivated by the above result, we propose a novel Stackelberg-game based double-auction mechanism where the link-supplier bids first, following which the users respond with their bids (Section~\ref{price_anticipation_delay_section}). The solution is characterized in terms of a \emph{Stackelberg equilibrium}. To show that the situation is now improved, we characterize the new efficiency loss in terms of the supplier's cost function, when the user pay-off functions are linear. For instance, for the quadratic link-cost function, the worst-case efficiency loss (with the worst-case taken over linear user pay-off functions) is at most 25\%.

\item We finally conduct a detailed numerical study to demonstrate the efficacy of the proposed Stackelberg-game based mechanism for a combination of link-cost functions (polynomial and exponential functions) and user utilities ($\alpha$-fair and $\log$ utilities)

\item We extend all of the above theoretical results to a setting with multiple links (see supplementary material).
\end{enumerate}

Our key contribution summarized in the third point above indicates that the situation, with the enforced timing of bids, is much improved. The neutral network-manager enforces the timing and enables the coordination that serves to protect the interests of system as a whole.

\vspace{-5mm}
\subsection{An Implementation Theory Perspective}
From an implementation theory perspective, the Iosifidis et al.\ \cite{iosifidis-etal15double-auction} mechanism in the price-taking scenario implements the social welfare maximization rule under the competitive equilibrium solution concept with the minimal message dimension of 1 (scalar signals). The above implementation ignores strategic behavior of individual agents. It is not possible to enforce such mechanisms in general because individual preferences may diverge from social welfare maximization. This is the price-anticipating scenario. It is anticipated that if we do not enlarge the signal space dimension there may be no mechanism, let alone the Iosifidis et al. mechanism, that can implement the social welfare maximization rule, under now the Nash equilibrium solution concept. This is why the price-anticipating scenario with non-strategic link suppliers suffered from an efficiency loss. What is surprising in our current setting is the dramatic increase in efficiency loss from at most 34\% (Johari et al. \cite{johari-etal2005efficiency}) to 100\% (contribution (2) of this paper). What is promising from our study is that this efficiency loss can be mitigated by structuring the interaction, by making the link player lead the interaction (contribution (3) of this paper). The solution concept is that of a Stackelberg equilibrium. Efficiency loss drops down to a value that depends on the supplier's cost function and is at most 25\% for quadratic costs and linear user pay-offs. This of course raises the question of what is the minimal signalling dimension in the price-anticipating scenario that implements the social welfare maximization rule in the Nash equilibrium solution concept. This is a very interesting question that is beyond the scope of this work. Our proposed scheme, which structures the interactions by asking the supplier to lead, reduces efficiency loss. It would be of utmost interest if this structuring also reduces the minimum signalling dimension for social welfare maximization in the Stackelberg equilibrium solution concept. We refer the reader to \cite{stoenescu2005decentralized} for an excellent discussion on the implementation theory perspective.

\subsection{Paper Outline}
The paper is organized as follows. In Section~\ref{system_model_section}, we propose our   system model, and discuss the problem formulation. The scenario with price-taking agents is discussed in  Section~\ref{price_taking_section}, while the price-anticipating agents' scenario is addressed in  Section~\ref{price_anticipating_section}. In Section~\ref{price_anticipation_delay_section}, we present our Stackelberg-game based mechanism, and characterize the worst-case efficiency loss for linear user pay-offs in terms of the link-supplier's cost function. Results from our numerical study are presented in Section~\ref{numerical_work_section}. To focus on the flow of key ideas, we have moved all the proofs to the Appendix. The paper concludes with some remarks in Section~\ref{conclusion_section}. Generalization of the above results to a setting comprising multiple link-suppliers are presented in the supplementary material (Appendix \ref{multiple_link_section} to \ref{ML_price_anticipation}).

\section{System Model}
\label{system_model_section}
We consider a model comprising  $M$ \emph{users} who intend to share the bandwidth available at a (single) \emph{link} of capacity $C>0$ owned by a \emph{link-supplier}. In the context of mobile-data offloading \cite{iosifidis-etal15double-auction}, the mobile-network operators and an access-point  (e.g., Wi-Fi, femtocell) operator constitute users and link-supplier, respectively, while the offloading capacity of the access-point can be regarded as the link. The mobile-network operators intend to buy a share of the limited bandwidth resource available at the access point to offload their macrocellular traffic, while the access point operator is interested in maximizing his profit. In the double auction terminology \cite{mcafee92dominant_strategy}, users are synonymous to buyers bidding for a share of a resource while the link-supplier is the seller. We refer to the users and the link-supplier collectively as \emph{agents}. The social planner, the entity that designs the mechanism (i.e., sets up the rules for information transfer, allocation, and payments) is referred to as the \emph{network-manager}.

Let $x_m$  denote the rate requested by user $m=1,2,\cdots,M$, and let  $y_m$  be the rate the {link-supplier} is willing to allocate to user $m$. Thus, $\textbf{x}=(x_1,x_2,\cdots,x_M)$ and $\textbf{y}=(y_1,y_2,\cdots,y_M)$ represent the \emph{rate-request} and \emph{rate-allocation} vectors, respectively. Let $y=\sum_{m} y_m$ denote the aggregate-rate allocated by the link-supplier to all users. For user $m$, the benefit of acquiring a rate of $x_m$ is represented by a pay-off function $U_m(x_m)$; we assume that $U_m$, $m=1,2,\cdots,M$, are concave, strictly increasing and continuously differentiable with finite $U_m'(0)$. Similarly, the cost incurred by the link-supplier for accepting to serve an aggregate rate of $y$ is given by $V(y)$, where $V$ is strictly convex, strictly increasing and continuously differentiable. Thus, the system optimal solution is the solution to the optimization problem:

SYSTEM
\begin{subequations}
\label{system_equn}
\begin{align}
\label{system_objective_equn}
\mbox{Maximize:}\hspace{2mm} & \vc{\sum}{}{m} U_m(x_m) - V\left(\vc{\sum}{}{m} y_m\right) \\
\label{system_capacity_constraint_equn}
\mbox{Subject to:}\hspace{2mm}  & \vc{\sum}{}{m}y_m\le C \\
\label{system_matching_constraint_equn}
& x_m\le y_m\ \forall m\\
\label{system_nonnegative_constraint}
&  x_m\ge0, y_m\ge0\ \forall m.
\end{align}
\end{subequations}

Continuity of the objective function and compactness of the constraint set imply that  an optimal solution $\textbf{x}^s=(x_1^s,x_2^s,\cdots,x_M^s)$ and $\textbf{y}^s=(y_1^s,y_2^s,\cdots,y_M^s)$ exists. Further,  if $U_m$ are strictly concave then (since $V$ is strictly convex) the solution is unique. Since $U_m$ are strictly increasing in $x_m$, an optimal solution must satisfy $\textbf{x}^s=\textbf{y}^s$. Thus, at optimality, the rate-requests (demand) and the rate-allocations (supply) are matched although the capacity $C$ may not be fully utilized.

 \begin{figure}[b]
 \vspace{-4mm}
 \centering
\includegraphics[scale=0.78]{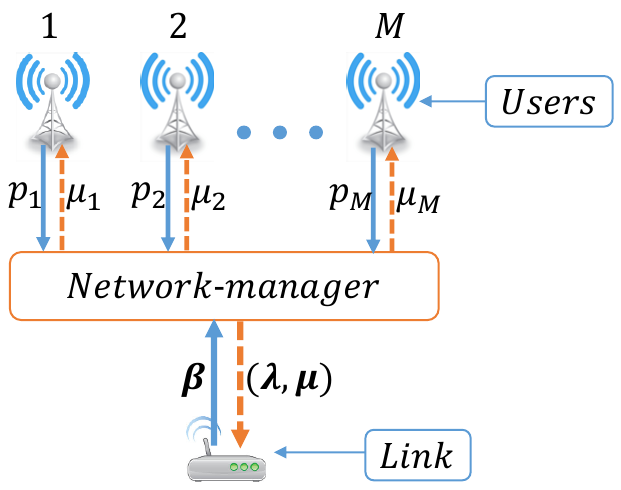}
 \caption{Direction of information exchange: solid lines represent the exchange of bids between the agents and the network-manager, while the dashed lines indicate the prices announced by the network-manager. \label{system_model_figure}}
 \end{figure}

A network-manager, however, cannot solve the formulation in (\ref{system_equn}) without the knowledge of user pay-offs and the link-cost function. Hence, we consider the following mechanism proposed by Iosifidis et al.\ in \cite{iosifidis-etal15double-auction} for rate allocation. Each user $m$ submits a \emph{bid} $p_m\ge0$ that denotes the amount he is willing to pay, while the link-supplier communicates signals $\beta_m$ $(m=1,2,\cdots,M)$ that implicitly indicate the amounts of bandwidth that he is willing to provide; we refer to $\textbf{p}:=(p_1,p_2,\cdots,p_M)$ and  $\bm{\beta}:=(\beta_1,\beta_2,\cdots,\beta_M)$ as the \emph{bids} submitted by the users and the link-supplier, respectively.

The network-manager is responsible for  fixing the \emph{prices} $\mu_m$ $(m=1,2,\cdots,M)$ and $\lambda$ that determines the rate allocation. The prices $\bm{\mu}:=(\mu_1,\mu_2,\cdots,\mu_M)$ and $\lambda$ are dual optimal of the following network problem proposed by Iosifidis et al.\ in \cite{iosifidis-etal15double-auction}:

NETWORK
\begin{subequations}
\label{network_equn}
\begin{align}
\label{net1_objective_equn}
\mbox{Maximize:}\hspace{2mm} &
\vc{\sum}{}{m}
~p_m \log(x_m) - \vc{\sum}{}{m}~\frac{y_m^2}{2\beta_m} \\
\label{net1_capacity_constraint_equn}
\mbox{Subject to:}\hspace{2mm}  & \vc{\sum}{}{m}y_m\le C \\
\label{net1_matching_constraint_equn}
& x_m\le y_m\ \forall m\\
\label{net1_nonnegative_constraint_equn}
&  x_m\ge0, y_m\ge0\  \forall m.
\end{align}
\end{subequations}
The above NETWORK problem is identical to the SYSTEM problem but with the true pay-off and cost functions replaced by {\em surrogate} pay-off and cost functions. See Fig.~\ref{system_model_figure} for an illustration of the direction of information exchange between the various entities of the model.

\emph{Remark:} In the NETWORK problem above we choose to use $\bm{\beta}$ instead of a related $\bm{\alpha}$ that was used in the original formulation by Iosifidis et al.\ in \cite{iosifidis-etal15double-auction}. The quantities $\bm{\alpha}$ and $\bm{\beta}$ are related by $\beta_m={1}/{\alpha_m}\ \forall m$. Thus,  $\beta_m$ are $\mathbb{R}_+$-valued, while each $\alpha_m$ is in general $\mathbb{R}_+ \cup \{+\infty\}$-valued.
Moreover, the signals in $\bm{\beta}$ are directly proportional to the amount of bandwidth the link-supplier is willing to share. For instance, a lower value of $\beta_m$ implies that the bandwidth shared by the link-supplier with user $m$ is low, and vice versa. In particular, $\beta_m=0$ implies that the link-supplier is unwilling to share any bandwidth with user $m$. This will be useful later while interpreting the Nash equilibrium bid-vectors (Theorem~\ref{pa_NE_theorem}).

In Section~\ref{price_taking_section} we first review the case when the users and the link-supplier are \emph{price-taking} whereby the prices set by the network-manager are simply accepted by all the agents without anticipating the effect that their bids  have  on the set prices. We then proceed to study the more-involved \emph{price-anticipating} scenario (Section~\ref{price_anticipating_section}). Here the agents recognize that the network-manager's prices are based on their bids. The agents hence anticipate the resulting allocation, payment, and eventually their pay-offs, and thus act accordingly. Our methodology in Sections~\ref{price_taking_section} and~\ref{price_anticipating_section} is similar to Johari et al. \cite{johari-etal2005efficiency}, but the outcome in the price-anticipating scenario is dramatically negative due to the presence of the strategic link-supplier, as we will soon see. We then propose a remedy via a Stackelberg framework (Section~\ref{price_anticipation_delay_section}) where the link-supplier is a lead player and the users are followers.

\section{Price-taking Scenario}
\label{price_taking_section}
The sequence of exchange of bids and prices (between the network-manager and the agents) that occur during the implementation of the  price-taking mechanism (PTM) is as shown in the following table.

\begin{tcolorbox}
{\small PRICE-TAKING MECHANISM (PTM)}
\begin{enumerate}
\item The network-manager  initiates the bidding process by fixing the prices $(\bm{\mu},\lambda)$.

\item The agents accept the prices and  respond by announcing their respective bids,  $\textbf{p}$ and $\bm{\beta}$ to the network-manager.

\item The network-manager allocates a rate of $x_m=p_m/\mu_m$ to user $m$ and receives a payment of $p_m$.  Simultaneously, the link-supplier is asked to supply a rate of  $y_m=\beta_m(\mu_m-\lambda)$ to user $m$; the total payment made by the network-manager to the link-supplier is  ${\sum}_{m} \beta_m{(\mu_m-\lambda)^2}$.
\end{enumerate}
\end{tcolorbox}

Given the prices $(\bm{\mu},\lambda)$ of the network-manager, the pay-off to user $m$, as a function of its bid $p_m$, is given by
\begin{eqnarray}
\label{pt_user_payoff_equn}
P_m\Big(p_m;\mu_m\Big)
= U_m\left(\frac{p_m}{\mu_m}\right) -p_m.
\end{eqnarray}
Naturally, user-$m$ will choose a $p_m$ that optimizes the above payoff function. Similarly, the pay-off to the link-supplier is given by
\begin{eqnarray}
\label{pt_link_payoff_equn}
\lefteqn{P_L\Big(\bm{\beta};(\bm{\mu},\lambda)\Big)} \nonumber  \\
& = & - V\left(\vc{\sum}{}{m} \beta_m\Big({\mu_m-\lambda}\Big)\right) + \vc{\sum}{}{m} \beta_m{\Big(\mu_m-\lambda\Big)^2}.
\end{eqnarray}
Thus, the bid $\bm{\beta}$ announced by the link-supplier is derived by optimizing the above payoff function.  

From the above expressions we see that the bids $(\textbf{p},\bm{\beta})$ that the agents announce explicitly depends on the prices $(\bm{\mu},\lambda)$ set by the network-manager. Thus, the network-manager has the authority to decide the operating point $(\textbf{p},\bm{\beta},\bm{\mu},\lambda)$ of the system. Being a neutral entity, the objective of the network-manager is naturally to   achieve social optimality (i.e., solution to the SYSTEM problem in (\ref{system_equn})). In the following (Theorem~\ref{competitive_equilibrium_theorem}) we prove that there exists prices $(\bm{\mu},\lambda)$ that will drive the system to a social optimal solution. Towards this end, we first introduce the notion of a \emph{competitive equilibrium} that encompasses the conditions necessary for optimality. In the following, unless mentioned otherwise, we assume that the agents\textquotesingle\ bids and the link-supplier\textquotesingle s prices are non-negative,  i.e., $p_m,\beta_m,\mu_m,\lambda\ge0\ \forall m$; also, we use $\textbf{0}$ to denote the vector of all-zeros of appropriate length.

\begin{definition}[Competitive Equilibrium \cite{johari-etal2005efficiency,johari-tsitsiklis04efficiency-loss}]
\label{CE_definition}
We say that $(\textbf{p},\bm{\beta},\lambda,\bm{\mu})$ constitutes a competitive equilibrium if the following conditions hold:
\begin{enumerate}
\item[(\bf{C1})]
$P_m(p_m;\mu_m) \ge P_m(\overline{p}_m;\mu_m)\ \forall \overline{p}_m\ge0, \forall m$\\
\item[(\bf{C2})] $P_L\Big(\bm{\beta};(\bm{\mu},\lambda)\Big) \ge P_L\Big(\overline{\bm{\beta}};(\bm{\mu},\lambda)\Big)\ \forall \overline{\bm{\beta}} \ge\textbf{0}$ \\
\item[(\bf{C3})] Define $\mathcal{M}=\Big\{m:\mu_m\ne\lambda\Big\}$ and
\begin{eqnarray}
\label{C0_equn}
\widehat{C}=\sqrt{  \left(\vc{\sum}{}{m} p_m \right) \left(\underset{m\in\mathcal{M}}{\sum}\beta_m \right) }.
\end{eqnarray}
Then, the following should hold:
 \begin{itemize}
 \item[(\bf{C3-a})]
 For all $m$,
\begin{eqnarray}
\label{C3_1_equn}
\frac{p_m}{\mu_m} = \beta_m\Big({\mu_m-\lambda}\Big);
\end{eqnarray}

 \item[(\bf{C3-b})]
 For all  $m\in\mathcal{M}$, the equality $\mu_m=\mu$ holds, where
 \begin{eqnarray}
 \label{mu1_equn}
\mu = {\vc{\sum}{}{i} p_i}\left/{\min\Big\{C,\widehat{C}\Big\}}\right.;
\end{eqnarray}

\item[(\bf{C3-c})] Furthermore,
 \begin{eqnarray}
 \label{lambda_equn}
 \lambda = \min\left\{0,\left(1-\left(\frac{C}{\widehat{C}}\right)^2\right) \frac{\vc{\sum}{}{i} p_i }{C}\right\}.
\end{eqnarray}
\end{itemize}
\end{enumerate}
\end{definition}

Condition (C1) above implies that the users do not benefit by deviating from their equilibrium bids $p_m$, when the prices $(\lambda,\bm{\mu})$ set by the network-manager are fixed. Similarly, (C2) implies that the link-supplier has no incentive to deviate from the equilibrium bid-vector $\bm{\beta}$. Although (C1) and (C2) result in the optimality of the users\textquotesingle\ and the link-supplier\textquotesingle s problem of maximizing their respective pay-offs, these conditions by themselves do not guarantee system-optimal performance. The conditions in (C3) (essentially derived from the optimality conditions for NETWORK) are crucial to guarantee that the prices $(\lambda,\bm{\mu})$ set by the network-manager are dual optimal for SYSTEM.  Condition (C3) along with (C1) and (C2)  can then be used to show the optimality of a competitive equilibrium. We summarize this result in the following theorem; in particular, we first prove the existence of a competitive equilibrium, and then derive its optimality property. This theorem is essentially an extension of the result due to Kelly \cite{kelly97charging-rate-control} and Kelly et al.\ \cite{kelly-etal98rate-control} (see also \cite{johari-etal2005efficiency} and \cite{johari-tsitsiklis04efficiency-loss}). The main difference that warrants an extension is the presence of the link-supplier as a strategic agent.

\begin{theorem}
\label{competitive_equilibrium_theorem}
When the agents are price-taking, there exists a competitive equilibrium, i.e., there exist  vectors $(\textbf{p},\bm{\beta},\lambda,\bm{\mu})$ satisfying (C1), (C2) and (C3). Moreover, given a competitive equilibrium $(\textbf{p},\bm{\beta},\lambda,\bm{\mu})$, the rate vectors $\textbf{x}$ and $\textbf{y}$ defined as $x_m={p_m}/{\mu_m}$ and $y_m=\beta_m(\mu_m-\lambda)$ ($\forall m$)  are optimal  for the problem SYSTEM in (\ref{system_equn}). 
\end{theorem}
\begin{IEEEproof}
The proof of the above theorem can be gleaned  from the results in  \cite{iosifidis-etal15double-auction} (although it is not explicitly stated in \cite{iosifidis-etal15double-auction}) where learning dynamics are used to show the convergence of the system to social optimality. For completeness, we however provide an alternative proof that is based on the direct approach of using Lagrangian techniques. Details are available in the supplementary material (Appendix \ref{app:proof-of-CE}).
\end{IEEEproof}

\vspace{-4mm}
\subsection*{Discussion on Practical Implementation}
Although the result in Theorem~\ref{competitive_equilibrium_theorem} suggests the existence of a competitive equilibrium, it is not immediately clear as to how the network-manager can determine such an equilibrium. Here we discuss an iterative algorithm  that can enable the network-manager to drive the system to an equilibrium. The algorithm is motivated by the best response dynamic and is an extension of PTM. The details are as follows.

The network-manager begins by fixing an initial price $(\bm{\mu}(0),\lambda(0))$. The agents simply accept the  prices and respond with their bids $(\textbf{p}(0),\bm{\beta}(0))$ (obtained by optimizing the respective pay-offs in (\ref{pt_user_payoff_equn}) and (\ref{pt_link_payoff_equn})). Given the bids $(\textbf{p}(0),\bm{\beta}(0))$, the network-manager sets new prices $(\bm{\mu}(1),\lambda(1))$ by solving the NETWORK problem in (\ref{network_equn}). The process continues until the values of prices and bids converge\footnote{Although convergence of such iterative algorithms is well known in the literature, a formal proof is beyond the scope of this work.}; essentially, steps 1 and 2 of PTM are iterated until the values converge. After convergence, step 3 is finally evoked to determine the rate allocation and the payments. In the above process, note that there is no direct exchange of information between the agents. All bids are collected by  the network-manager, which then communicates only the required prices to the respective agents (i.e., only $(\mu_m,\lambda)$ is communicated to user-$m$, while the entire $(\bm{\mu},\lambda)$ is given to the link-supplier).

\section{Price-anticipating Scenario -- Nash Formulation}
\label{price_anticipating_section}
In the price-anticipating scenarios  agents initiate the bidding process, which is in contrast to that in the price-taking scenario. Further, in the Nash formulation the agents simultaneously announce their bids. Specifically, the sequence of exchanges that take place in the price-anticipating mechanism of the Nash formulation (PAM-N) is as given below.

\begin{tcolorbox}
{\small PRICE ANTICIPATING MECHANISM -- NASH (PAM-N)}
\begin{enumerate}
\item Agents  initiate the bidding process by \emph{simultaneously} announcing their bids $\textbf{p}$ and $\bm{\beta}$ to the network-manager.

\item The network-manager then computes the prices $(\bm{\mu}(\textbf{p},\bm{\beta}),\lambda(\textbf{p},\bm{\beta}))$ where $\bm{\mu}(\textbf{p},\bm{\beta})$ is the vector $(\mu_1(\textbf{p},\bm{\beta}),\cdots,\mu_M(\textbf{p},\bm{\beta}))$. The above prices are dual optimal for the NETWORK problem in (\ref{network_equn}).

\item The payments and the allocated rates are exactly as in the price-taking mechanism, but with $(\bm{\mu},\lambda)$ replaced by $(\bm{\mu}(\textbf{p},\bm{\beta}),\lambda(\textbf{p},\bm{\beta}))$.
\end{enumerate}
\end{tcolorbox}

Given the agents' bids $(\textbf{p},\bm{\beta})$, it is natural for the network-manager to go ahead and solve the NETWORK problem in (\ref{network_equn}).  The dual optimal variables, denoted $(\bm{\mu}(\textbf{p},\bm{\beta}),\lambda(\textbf{p},\bm{\beta}))$,  are then used by the network-manager to determine the amount of rate and incentives to be exchanged. An explicit expression for the above prices is presented in the following lemma.

\begin{lemma}
\label{lagrange_multipliers_lemma}
Given any vector $(\textbf{p},\bm{\beta})$ of users\textquotesingle\ and link-supplier\textquotesingle s bids, the prices $(\lambda(\textbf{p},\bm{\beta}),\bm{\mu}(\textbf{p},\bm{\beta}))$ set by the network-manager are given by
\begin{eqnarray}
\label{lbd_equn}
\lambda(\textbf{p},\bm{\beta})
&=&
\left\{
\begin{array}{ll}
0 & \mbox{ if } \vc{\sum}{}{i}\sqrt{{p_i}{\beta_i}}\le C \\
f^{-1}_{\textbf{p},\bm{\beta}}(C) & \mbox{ otherwise, }
\end{array}
\right.
\end{eqnarray}
where $f^{-1}_{\textbf{p},\bm{\beta}}$ is the inverse of $f_{\textbf{p},\bm{\beta}}$ defined as
\begin{eqnarray}
\label{f_func_equn}
f_{\textbf{p},\bm{\beta}}(t)
&=& \vc{\sum}{}{i} \left(\frac{2p_i}{t+\sqrt{t^2+4\frac{p_i}{\beta_i}}}\right),
\end{eqnarray}
and for $m=1,2,\cdots,M$
\begin{eqnarray}
\label{mu_equn}
\mu_m(\textbf{p},\bm{\beta})
&=& \frac{\lambda(\textbf{p},\bm{\beta}) + \sqrt{\lambda(\textbf{p},\bm{\beta})^2+4\frac{p_m}{\beta_m}}}{2}.
\end{eqnarray}
\end{lemma}
\begin{IEEEproof}
See Appendix~\ref{lagrange_multipliers_lemma_appendix}.
\end{IEEEproof}

Using the above results in  (\ref{pt_user_payoff_equn}), the pay-off that user-$m$ ($m=1,2,\cdots,M$) would achieve can be expressed as follows  (where, for simplicity, we have used $\lambda$ for $\lambda(\mathbf{p},\bm{\beta})$):
\begin{eqnarray}
\label{user_payoff_equn}
\lefteqn{Q_m(p_m,\mathbf{p}_{-m},\bm{\beta})
=U_m\left(\frac{p_m}{\mu_m(\textbf{p},\bm{\beta})}\right) - p_m} \nonumber\\
&&=
\left\{
\begin{array}{ll}
U_m\left(\sqrt{{p_m}{\beta_m}}\right) - p_m \mbox{ if } \vc{\sum}{}{i}\sqrt{{p_i}{\beta_i}}\le C \\
U_m\left(\frac{2p_m}{\lambda+\sqrt{\lambda^2+4\frac{p_m}{\beta_m}}}\right) - p_m  \mbox{ otherwise},
\end{array}
\right.
\end{eqnarray}
where $\mathbf{p}_{-m}=(p_1,\cdots,p_{m-1},p_{m+1},\cdots,p_M)$ denotes the bids of all users other than user-$m$. Similarly, for the link-supplier we have
\begin{eqnarray}
\lefteqn{Q_L(\bm{\beta},\mathbf{p})}\nonumber \\
\label{link_payoff_equn}
&=&
\left\{
\begin{array}{ll}
-V\left(\vc{\sum}{}{m}\sqrt{{p_m}{\beta_m}}\right) + \underset{m}{\sum} p_m \hspace{2mm} \mbox{ if } \vc{\sum}{}{i}\sqrt{{p_i}{\beta_i}}\le C \\
-V(C) +  \underset{m}{\sum} \frac{1}{\beta_m} \left(\frac{2p_m}{\lambda+\sqrt{\lambda^2+4\frac{p_m}{\beta_m}}}\right)^2  \mbox{ otherwise}.
\end{array}
\right.
\end{eqnarray}
The quantity $V(C)$ in the above expression is due to complementary slackness conditions which imply
\begin{eqnarray*}
\underset{m}{\sum}\frac{p_m}{\mu_m(\mathbf{p},\bm{\beta})}=\underset{m}{\sum}y_m=C \mbox{ whenever } \lambda>0.
\end{eqnarray*}

The users and the link-supplier recognize that their bids affect the prices and the allocation. Acting as rational and strategic agents, they now anticipate these prices. The appropriate notion of an equilibrium in this context is the following.

\begin{definition}[Nash Equilibrium]
\label{NE_definition}
A bid vector  $(\textbf{p},\bm{\beta})$ is a Nash equilibrium if, for all $m=1,2,\cdots,M$, we have
\begin{eqnarray*}
Q_m(p_m,\mathbf{p}_{-m},\bm{\beta}) &\ge& Q_m(\overline{p}_m,\mathbf{p}_{-m},\bm{\beta})\ \forall \overline{p}_m\ge0 \\
Q_L(\bm{\beta},\mathbf{p})&\ge& Q_L(\overline{\bm{\beta}},\mathbf{p})\ \forall \overline{\bm{\beta}}\ge\textbf{0}.
\end{eqnarray*}
\QED
\end{definition}

\noindent
When $\sum_{i}\sqrt{{p_i}{\beta_i}} < C$, the link is not fully utilized. In this case the Lagrange multiplier $\lambda = \lambda(\textbf{p},\bm{\beta}) = 0$. Examination of (\ref{user_payoff_equn}) and (\ref{link_payoff_equn}) indicates that the payments made by the users are all passed on to the link-supplier. This may be interpreted as follows: for a given  set of payments, the link-supplier bids are such that the link is viewed as a costly resource and the network-manager passes on all his revenue to the link-supplier. The link-supplier is thus assured of this revenue even if his link is not fully utilized. If, on the other hand, the link-supplier\textquotesingle s bids are such that $\sum_{i}\sqrt{{p_i}{\beta_i}} > C$, then $\lambda > 0$, and it is clear from (\ref{link_payoff_equn}) that not all the collected revenue is passed on to the link-supplier. Indeed, since $\lambda > 0$, we have
\vspace{-2.5mm}
\begin{eqnarray*}
\sum_m \frac{1}{\beta_m} \left(\frac{2p_m}{\lambda+\sqrt{\lambda^2+4\frac{p_m}{\beta_m}}}\right)^2 <\sum_m p_m
\end{eqnarray*}
where the right-hand side is obtained when $\lambda = 0$. The actions of the link-supplier as a strategic agent creates a situation of conflict and results in the following undesirable equilibrium.

\begin{theorem}
\label{pa_NE_theorem}
When the users and the link-supplier are price-anticipating, the only Nash equilibrium is $(\textbf{p}^o,\bm{\beta}^o)$ where $p_m^o=0$ and $\beta_m^o=0$ for all $m=1,2,\cdots,M$.
\end{theorem}
\begin{IEEEproof}
See Appendix~\ref{pa_NE_theorem_appendix}.
\end{IEEEproof}

\noindent
Thus, in the price-anticipating setting, efficiency loss is 100\%, which we interpret as a market break-down. Indeed, at $\bm{\beta}^o=\textbf{0}$, the link-supplier is assured an income of $\sum_m p_m$. Given this guaranteed income, he minimizes his cost by supplying zero capacity. The resulting equilibrium is one with the lowest efficiency, and the situation is vastly different from the setting when the link-supplier is not viewed as an agent \cite{johari-etal2005efficiency}.

\vspace{-2mm}
\subsection*{Discussion on Practical Implementation}
Although it is completely not favorable to operate at the Nash equilibrium $(\textbf{p}^o,\bm{\beta}^o)$,  for completeness we still present an iterative algorithm (like in the price-taking scenario) that would converge to $(\textbf{p}^o,\bm{\beta}^o)$. This discussion will also enable us to reinforce the differences between the price-taking and the price-anticipating scenarios. Further, knowing the subtle  variations between the Nash-equilibrium and the upcoming Stackelberg-equilibrium convergent algorithms will enable one to be cautious while implementation. The details of the algorithm are as follows.

As in the price-taking scenario, the algorithm that we present is essentially an iterative version of PAM-N mechanism. For simplicity, we assume $C=\infty$ so that the results in Lemma~\ref{lagrange_multipliers_lemma} reduces to $\lambda:=\lambda(\textbf{p},\bm{\beta})=0$ and $\mu_m:=\mu_m(\textbf{p},\bm{\beta})=\sqrt{p_m/\beta_m}$; the finite $C$ case can be similarly handled. Now, the algorithm begins with the agents submitting their initial bids, denoted $(\textbf{p}(0),\bm{\beta}(0))$. The network-manager computes the prices $\bm{\mu}(0)=(\mu_1(0),\cdots,\mu_M(0))$, and communicates the appropriate prices to the agents.  Receiving $\mu_m(0)$, user-$m$ computes $\beta_m(0) = p_m(0)/\mu_m(0)^2$, using which in (\ref{user_payoff_equn}) user-$m$ determines $p_m(1)$ that optimizes its pay-off. Simultaneously, given $\bm{\mu}(0)$, link-supplier computes $p_m(0)=\mu_m(0)^2 \beta_m(0)$ (for $m=1,2,\cdots,M$), and then determines $\bm{\beta}(1)$ that optimizes the pay-off in (\ref{link_payoff_equn}). The agents (separately but simultaneously) submit the new bids $(\textbf{p}(1),\bm{\beta}(1))$, and the process continues until the values of the bids and the prices converge (i.e., steps 1 and 2 of PAM-N are iterated until convergence). After convergence, the rate allocation and payments are processed as per step 3.

\section{Price-anticipating Scenario -- Stackelberg Formulation}
\label{price_anticipation_delay_section}

\begin{tcolorbox}
{\small PRICE-ANTICIPATING MECHANISM -- STACKELBERG  (PAM-S)}
\begin{enumerate}
\item The link-supplier first announces his bid $\bm{\beta}$ to the network-manager. The network-manager makes this information  available to all users.

\item The users then send their bids $p_m^{\bm{\beta}}$ (that potentially depends on the value of $\bm{\beta}$) to the network-manager. Let $\bm{p}^{\bm{\beta}}=(p_1^{\bm{\beta}},p_2^{\bm{\beta}},\cdots,p_M^{\bm{\beta}})$.

\item The network-manager then computes the prices $(\bm{\mu}(\textbf{p}^{\bm{\beta}},\bm{\beta}),\lambda(\textbf{p}^{\bm{\beta}},\bm{\beta}))$ by solving the NETWORK problem in (\ref{network_equn}).

\item  The payments and the rates-allocated are exactly as in the price-taking mechanism, but with $(\bm{\mu},\lambda)$ replaced by $(\bm{\mu}(\textbf{p}^{\bm{\beta}},\bm{\beta}),\lambda(\textbf{p}^{\bm{\beta}},\bm{\beta}))$.
\end{enumerate}
\end{tcolorbox}

In view of the break-down of the market when both the users and the link-supplier are simultaneously price-anticipating, we design an alternative scheme that is based on a Stackelberg game formulation. The sequence of exchanges in our new mechanism is as given in the table above.

Thus, in the Stackelberg-game modification to the price-anticipating scenario, the network-manager collects bids from the agents in stages -- the link-supplier leads and the users follow. The analysis of this mechanism proceeds as follows. Given a $(\bm{\beta},\bm{p})$, the expression for the prices set by the network-manager are as in Lemma~\ref{lagrange_multipliers_lemma}. As a result, the expressions for the users\textquotesingle\  and the link-supplier\textquotesingle s pay-off functions are exactly as in (\ref{user_payoff_equn}) and (\ref{link_payoff_equn}), respectively, but with $\textbf{p}$ replaced by $\textbf{p}^{\bm{\beta}}$. Using these pay-off functions, we characterize the solution in terms of Stackelberg equilibrium:

\begin{definition}[Stackelberg Equilibrium]
A bid vector $(\bm{\beta},\textbf{p}^{\bm{\beta}})$ is a Stackelberg equilibrium if, for all $m=1,2,\cdots, M$, we have
\begin{eqnarray*}
Q_m(p_m^{\bm{\beta}},\mathbf{p_{-m}^{\bm{\beta}}},\bm{\beta}) &\ge& Q_m(\overline{p}_m,\mathbf{p_{-m}^{\bm{\beta}}},\bm{\beta})\hspace{3mm} \forall \overline{p}_m\ge0 \\
Q_L(\bm{\beta},\mathbf{p}^{\bm{\beta}})&\ge& Q_L(\overline{\bm{\beta}},\mathbf{p}^{\bm{\overline{\beta}}})\hspace{3mm} \forall \overline{\bm{\beta}}\ge\textbf{0}.
\end{eqnarray*}
\QED
\end{definition}
Observe that the bid-vector $\bm{\beta}$ announced by the link-supplier in step-1 anticipates the user bids $\textbf{p}^{\bm{\beta}}$ of step-2. For a given $\bm{\beta}$, the bids submitted by the users is in anticipation of the prices the network-manager announces in step-3.

For the ease of exposition, we assume that $C=\infty$  so that the capacity constraint is not binding (the case where $C$ is finite can be similarly handled). Thus, recalling (\ref{lbd_equn}) and (\ref{mu_equn}), we have  $\lambda(\textbf{p},\bm{\beta})=0$ and $\mu_m(\textbf{p},\bm{\beta})=\sqrt{\frac{p_m}{\beta_m}}$. As a result the pay-off functions in (\ref{user_payoff_equn}) and (\ref{link_payoff_equn}) can be simply expressed as
\begin{eqnarray}
\label{user_payoff_simple_equn}
Q_m(p_m,\mathbf{p}_{-m},\bm{\beta}) &=& U_m\left(\sqrt{{p_m}{\beta_m}}\right) - p_m \\
\label{link_payoff_simple_equn}
Q_L(\bm{\beta},\mathbf{p}) &=& -V\left(\vc{\sum}{}{m}\sqrt{{p_m}{\beta_m}}\right) + \underset{m}{\sum} p_m. \quad \quad
\end{eqnarray}
This simplification will enable us to focus on the key ideas rather than dwell on the technicalities arising from a finite $C$ (which can although be handled but is cumbersome and not very enlightening).

From (\ref{user_payoff_simple_equn}) we see that the user pay-offs are independent of the bids submitted by the other users. As a result, for a given $\bm{\beta}$,  the unique equilibrium strategy for user-$m$ is given by
\begin{eqnarray}
\label{pmbeta_equn}
p_m^{\bm{\beta}} &=&
\argmax_{p_m\ge0}\ \Big(U_m\left(\sqrt{{p_m}{\beta_m}}\right) - p_m\Big).
\end{eqnarray}
In Lemma~\ref{rbeta_lemma} we report the expression for $p_m^{\bm{\beta}}$ that is  obtained by solving (\ref{pmbeta_equn}).
\begin{lemma}
\label{rbeta_lemma}
For a given $\bm{\beta}$ we have
\begin{eqnarray}
\label{eqn:pmbeta}
p_m^{\bm{\beta}} =
\left\{
\begin{array}{ll}
\frac{r_{\beta_m}^2}{\beta_m} & \mbox{ if } \beta_m > 0 \\
0 & \mbox{ otherwise}
\end{array}
\right.
\end{eqnarray}
where $r_{\beta_m}$ is the fixed point of $U'_m(r)={2r}/{\beta_m}$.
\end{lemma}
\begin{IEEEproof}
See Appendix~\ref{rbeta_lemma_appendix}.
\end{IEEEproof}

We extend the definition of $r_{\beta_m}$ in the above lemma by defining $r_{\beta_m}=0$ if $\beta_m=0$.  It is then easy to see that $r_{\beta_m} = \sqrt{p_m^{\bm{\beta}} \beta_m }$ is the allocation to user $m$. Plugging the above result into (\ref{link_payoff_simple_equn}), we compute the optimal $\bm{\beta}$ that the link-supplier should announce in step-1 as
\begin{eqnarray}
\label{beta_star_equn}
{\bm{\beta}}^* \in \mathcal{B}^*= \argmax_{\bm{\beta} \geq \bm{0}} \left\{-V\left(\sum_m r_{\beta_m}\right) +\sum_m \frac{r_{\beta_m}^2}{\beta_m}\right\},
\end{eqnarray}
where $\bm{\beta} \geq \bm{0}$ means component-wise inequality.

For any ${\bm{\beta}}^* \in \mathcal{B}^*$ it is clear that $(\bm{\beta^*},\bm{p}^{\bm{\beta^*}})$ constitutes a Stackelberg equilibrium, where the rate allocated to user-$m$  is given by $x_m^{\bm{\beta^*}}=y_m^{\bm{\beta^*}}=\sqrt{p_m^{\bm{\beta^*}}\beta_m^*}=r_{\beta_m^*}$. However, we first need to assert the existence of a solution  ${\bm{\beta}}^*$, i.e., that the set $\mathcal{B}^*$ is nonempty.

\begin{theorem}
\label{stackelberg-existence-lemma}
Suppose $U_m(\cdot)$ and $V(\cdot)$  satisfy the following: $xU_m'(x)\rightarrow\infty$ and ${V(x)}/{x}\rightarrow\infty$   as $x\rightarrow\infty$. Then the set $\mathcal{B}^*$ is nonempty. Hence, under the above assumptions on the pay-offs and cost function, a Stackelberg equilibrium exists.
\end{theorem}
\begin{IEEEproof}
See Appendix~\ref{stackelberg-existence-lemma_appendix}.
\end{IEEEproof}

\noindent
\emph{Remark:} We would like to emphasize that the above assumptions on  utility and cost functions are not very restrictive. Specifically, the condition imposed on the utility function is satisfied by $\alpha$-fair utility functions $U_m(x)=x^{1-\alpha}/(1-\alpha)$ for the range of  $\alpha\in(0,1)$. Although the popular utility function $\log(1+x)$ is ruled out, the functions $\log(1+x^{q})$ for  $q\in(0,1)$ are allowed.  Similarly, the condition on the link-cost function is satisfied by a wide range of convex functions including polynomials, exponentials etc.; only those that are asymptotically linear are ruled out. Finally, we also note that it is not possible to assert the uniqueness of $\beta^*$. The main challenge that limits us from proving (or disproving) this result lies in the intricate form of the  function $r_{\beta_m}/\beta_m$ whose concavity (or non-concavity) property is not easy to establish (although it can be shown that $r_{\beta_m}$ is increasing in $\beta_m$).

\subsection*{Discussion on Practical Implementation}
The algorithm begins with the link-manager first announcing its initial bid $\bm{\beta}(0)$ to the network-manager. The network-manager communicates $\beta_m(0)$ to user-$m$ ($m=1,2,\cdots,M$) 
All users announce their respective prices $p_m(0)=p_m^{\bm{\beta}(0)}$ by solving (\ref{eqn:pmbeta}). The network-manager computes prices ${\mu}_m(0)=\sqrt{p_m(0)/\beta_m(0)}$, and communicates the price vector $\bm{\mu}(0)$  to the link-supplier.  The link-supplier, extracting  $\bm{p}(0)$ from the received price information, will evaluate its pay-off using (\ref{link_payoff_simple_equn}). Now, unlike the  Nash-equilibrium algorithm which determines $\bm{\beta}(1)$ by simply optimizing (\ref{link_payoff_simple_equn}),   here the link-supplier's objective is instead to find a $\bm{\beta}(1)$ that is in the direction of solving the problem in (\ref{beta_star_equn}). However, since  $r_{\beta_m}$ as a function of $\beta_m$ is unknown to the link-supplier, the problem of optimizing (\ref{beta_star_equn}) falls within the realm of \emph{blackbox optimization}. The link-supplier is hence expected to use a stochastic search algorithm such as simulated annealing, hill climbing, etc., to iteratively approach a Stackelberg equilibrium $\beta^*$ in (\ref{beta_star_equn}). Once the link-supplier's search algorithm converges, the network-manager announces the final rates and payments.

\subsection{Stackelberg Equilibrium for Linear User Pay-offs}
In the remainder of this section, we restrict our attention to linear user pay-offs. In this case an explicit expression for the Stackelberg equilibrium can be derived. Specifically, suppose that the user pay-offs are of the form $U_m(x_m)=c_mx_m$ where $c_m>0$ ($m=1,2,\cdots,M$).  Without loss of generality, assume that $c_1=\max_m\{c_m\}$. The Stackelberg equilibrium can then be computed as follows.

First, fix a $\bm{\beta}$. Recalling  Lemma~\ref{rbeta_lemma}, we have
\begin{eqnarray*}
r_{\beta_m}=\frac{\beta_m{U}'_m(r_{\beta_m})}{2}  = \frac{\beta_m c_m}{2}
\end{eqnarray*}
so that the equilibrium bid of user-$m$ can be written as
\begin{eqnarray}
{p}_m^{\bm{\beta}}=\frac{r_{\beta_m}^2}{\beta_m} = \frac{\beta_m c_m^2}{4}.
\end{eqnarray}
Substituting for $r_{\beta_m}$  in (\ref{beta_star_equn}), we obtain
\begin{eqnarray}
\label{beta_star_equn_R1}
{\bm{\beta}}^* \in \underset{\bm{\beta} \geq \bm{0}}{\argmax}\left\{  -V\left(\sum_m \frac{\beta_m c_m}{2}\right) +\sum_m \frac{\beta_m c_m^2}{4}
\right\}.
\end{eqnarray}
Since the link-cost function $V(\cdot)$ is strictly convex, it follows that there is a unique $\beta^*$ that solves the above problem. The explicit expression of $\beta^*$ is reported in the following lemma.

\begin{lemma}
\label{beta_solution_lemma}
Suppose that $v(x):=V'(x)$ satisfies $v(0)=0$. Then, the solution to the problem in (\ref{beta_star_equn_R1}) is given by $\bfb^*=(\beta_1^*,\beta_2^*,\cdots,\beta_M^*)$ where
\begin{eqnarray}
\label{beta_m_star_equn_R1}
\beta_m^* =
\left\{
\begin{array}{ll}
 \frac{2}{c_1}v^{-1}(\frac{c_1}{2}) & \mbox{ if } m=1 \\
 0 & \mbox{ otherwise }
 \end{array}
 \right.
\end{eqnarray}
\end{lemma}
\begin{IEEEproof}
Available in Appendix~\ref{beta_solution_lemma_appendix}
\end{IEEEproof}

\noindent
\emph{Remark:} The condition $v(0)=0$ imposed on the derivative of the cost function $V(\cdot)$ is necessary to  ensure that $\beta_1^*$ in (\ref{beta_m_star_equn_R1}) is positive. Thus, the market is guaranteed to be active for any set of linear user utilities. The above condition is not very restrictive as it is satisfied by the polynomial cost functions (i.e., $V(x)=a x^n$). The following modification of the exponential cost function also satisfies the above property: $V(x)=e^{ax}-(ax+1)$ for $a\ge 1$.

The equilibrium bids of users in response to this optimized $\bfb^*$ is then given by
\begin{eqnarray}
\label{stackelberg_unfair_allocation_equn}
p_m^{\bfb^*} =
\left\{
\begin{array}{ll}
\frac{c_1}{2} v^{-1}\left(\frac{c_1}{2}\right) & \mbox{ if } m=1 \\
0 & \mbox{ otherwise.}
\end{array}
\right.
\end{eqnarray}
Thus, when the user pay-offs are linear, the link-supplier allocates all the bandwidth to the ``best" user (i.e., the one with the maximum slope $c_m$); in return, the best user  alone makes a positive payment to the link-supplier.

The rate allocated to user $m$ at equilibrium is
\begin{eqnarray}
\label{xm_eq_rm_equn}
x_m^{\bfb^*} &=&  r_{\beta_m^*} \nonumber \\
&=& \sqrt{p_m^{\bm{\beta^*}}\beta_m^*} \nonumber\\
&=&
\left\{
\begin{array}{ll}
 v^{-1}\left(\frac{c_1}{2}\right) & \mbox{ if } m=1 \\
0 & \mbox{ otherwise.}
\end{array}
\right.
\end{eqnarray}
The total rate served by the link-supplier at equilibrium is given by  $\sum_m y_m^{\bm{\beta^*}}=\sum_m x_m^{\bm{\beta^*}}=v^{-1}\left(\frac{c_1}{2}\right)$.

\subsection{Lower Bound on Efficiency} 
Given a Stackelberg equilibrium $(\bm{\beta^*},\bm{p}^{\bm{\beta^*}})$ the \emph{efficiency} is defined as the ratio of the utility at equilibrium (\emph{Stackelberg utility}) to the system optimum (\emph{social utility}):
\begin{eqnarray}
\label{PoA_equn}
\mathcal{E}(\{U_m\};V) = \frac{\sum_m U_m\left(x_m^{\bm{\beta^*}}\right) - V\left(\sum_m x_m^{\bm{\beta^*}}\right)}{\sum_m U_m\left(x_m^{s}\right) - V\left(\sum_m x_m^{s}\right)}
\end{eqnarray}
where $x_m^s$ denotes the social optimum allocation to user $m$ (obtained by solving SYSTEM in (\ref{system_equn})). Note that we have emphasized the dependency of efficiency on $(\{U_m\};V)$ by incorporating these into the notation for efficiency.

When the link-supplier is non-strategic, from Johari et al.\ \cite{johari-etal2005efficiency} it is known that the bound on efficiency is $(4\sqrt{2}-5)$, i.e., $\mathcal{E}(\{U_m\},V)\ge(4\sqrt{2}-5)$ for any general collection of user pay-off functions $\{U_m\}$ (the loss in efficiency is thus no more than $34\%$). The above bound is obtained in \cite{johari-etal2005efficiency} by doing the following.
\begin{itemize}
\item[(a)] Show that the users' equilibrium bids in the \emph{original game} (with general user pay-off functions) constitutes an equilibrium in an \emph{alternate game} with appropriately chosen linear pay-off functions.

\item[(b)] Use this to show that the efficiency in the original game is bounded below by the efficiency achieved in the alternate game.

\item[(c)] Finally, minimize the efficiency over the set of all linear pay-offs; this can be explicitly computed and is $(4\sqrt{2}-5)$.
\end{itemize}

In our case, although (a) holds\footnote{Formally, we can show that for any given $\bm{\beta}$, the equilibrium strategy $\bm{p}^{\bm{\beta}}_m$ for the users in the original game with pay-off functions $\{U_m\}$ is also an equilibrium strategy for the users in an alternate game with linear pay-offs $\{\overline{U}_m\}$, where $\overline{U}_m(x_m)=c_m x_m$ with $c_m=U_m'(r_{\beta_m^*})$.} for any given $\bfb$, there is a subtle issue\footnote{Our conference version \cite{naveen2018double} missed this subtle point and incorrectly made a more general claim that the lower bound held for a larger class of user pay-offs.}. Since the link-supplier is also strategic, the original game and the alternate game (with linear user pay-offs) may not have identical Stackelberg equilibria. In particular, the $\bfb^*$ that optimizes the objective in (\ref{beta_star_equn}) may not necessarily optimize
\begin{eqnarray}
\max_{\bm{\beta} \geq \bm{0}} \left\{-V\left(\sum_m\frac{\beta_m a_m}{2}\right) +\sum_m \frac{\beta_m a_m^2}{4}\right\},
\end{eqnarray}
which is the objective corresponding to the game with linear pay-offs: $\overline{U}_m(x_m)=a_m x_m$ with $a_m=U_m'(r_{\beta_m^*})$. Thus, (a) and (b) may not hold for general user pay-offs. However, an analog of (c) continues to hold if we restrict our attention to the ensemble of all linear user pay-offs. The lower bound on efficiency will however depend on the link-suppliers cost function $V(x)$. This result is detailed in the following theorem.

\begin{figure*}[t]
\centering
\subfigure[]{
\includegraphics[scale=0.36]{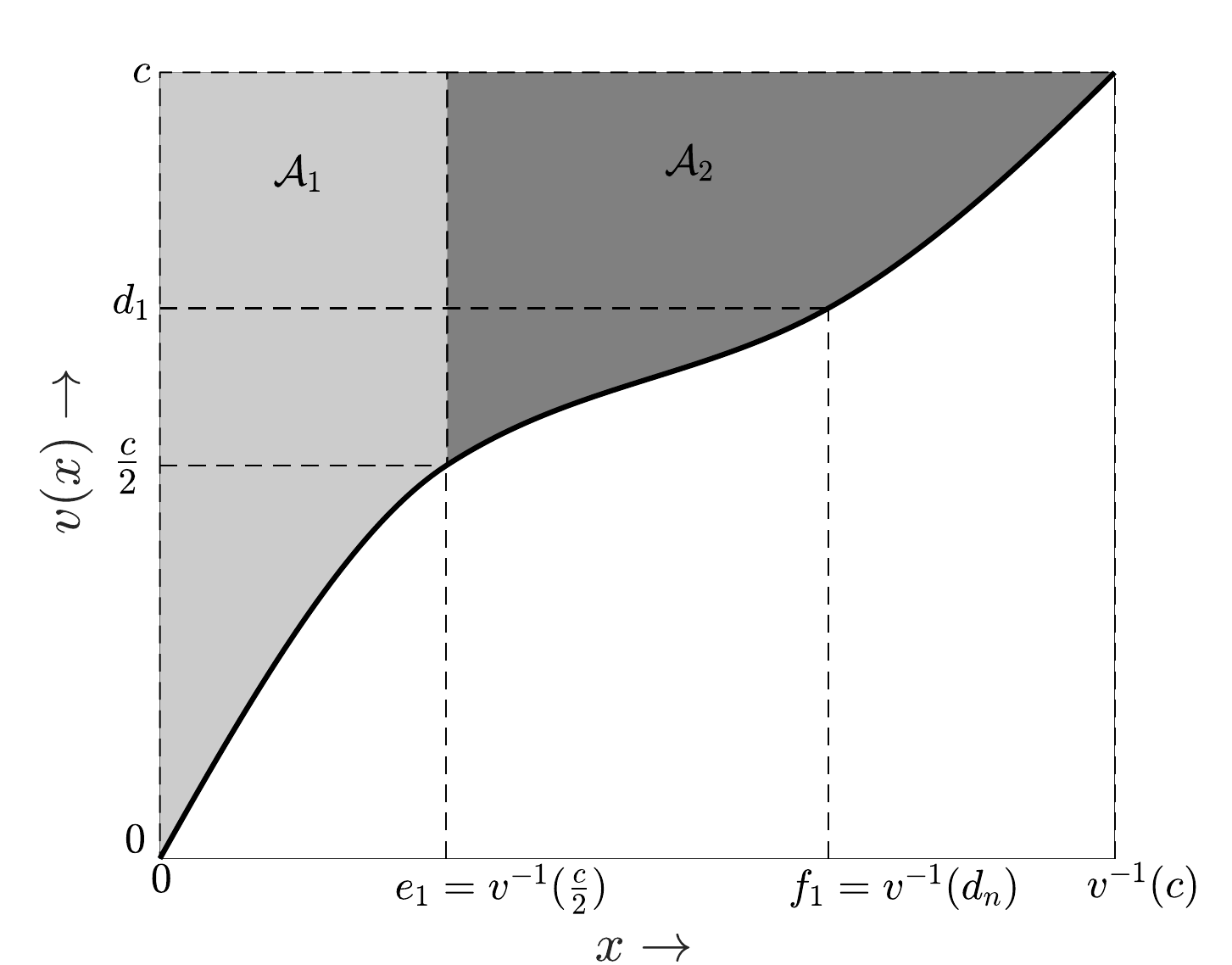}
        \label{geo_int_figure}
}
\hspace{-4mm}
\subfigure[]{
\includegraphics[scale=0.36]{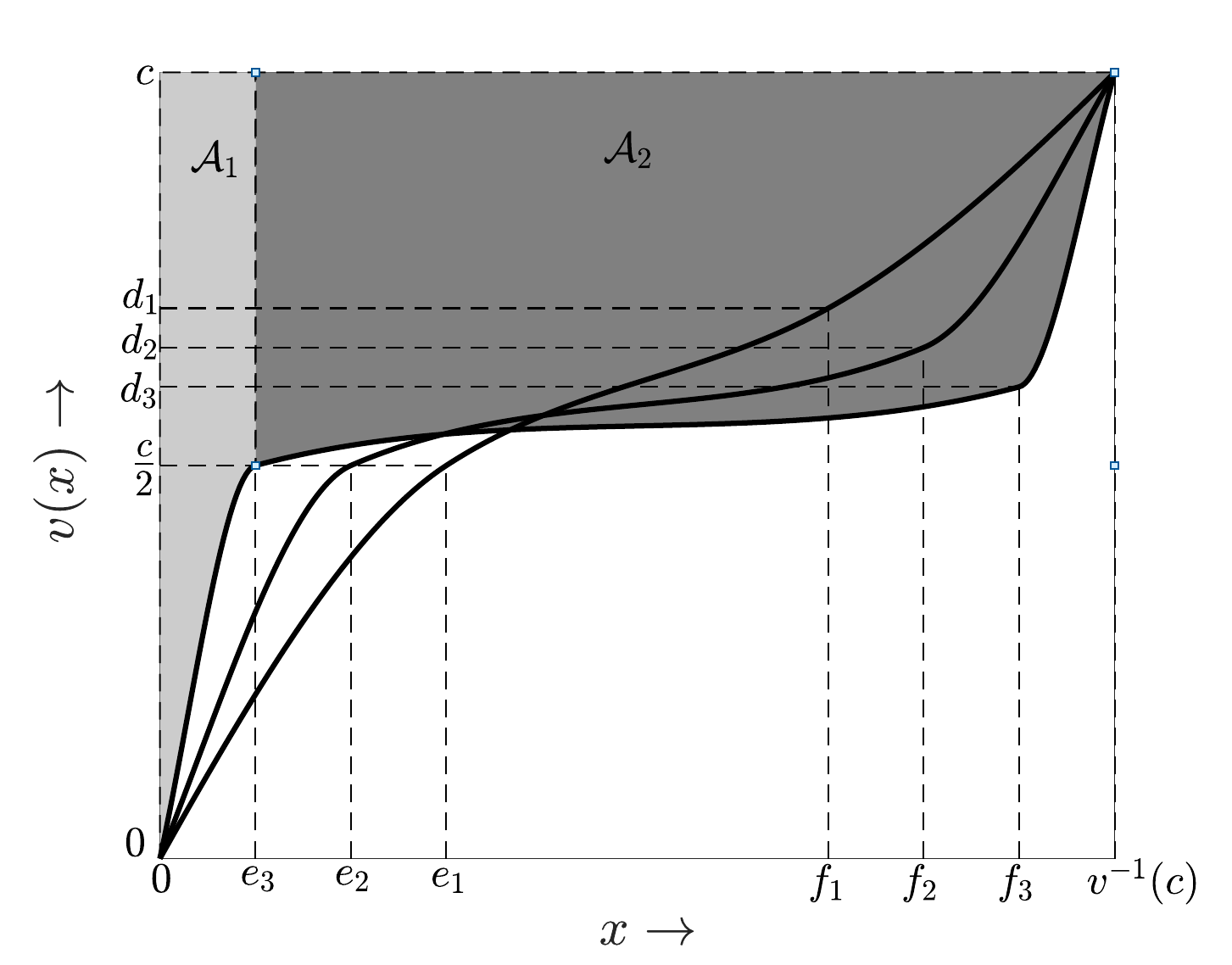}
        \label{different_plots_figure}
}
\caption{\subref{geo_int_figure} Geometric interpretation of the efficiency bound; \subref{different_plots_figure} Construction of a sequence of degenerate link-cost functions whose efficiency decreases to $0$.
}
\vspace{-5mm}
\end{figure*}

\begin{theorem}
\label{poa_theorem}
Fix a link-cost function $V(\cdot)$ that satisfies $v(0)=0$ where $v(\cdot):=V'(\cdot)$. Then, for any collection of linear user pay-offs $\{U_m\}$, we have
\begin{eqnarray}
\label{poa_bound_equn}
\mathcal{E}(\{U_m\};V) &\ge& \inf_{c>0}\ \frac{c v^{-1}(\frac{c}{2})-V(v^{-1}(\frac{c}{2}))}{c v^{-1}(c)-V(v^{-1}(c))}
\end{eqnarray}
\end{theorem}
\begin{IEEEproof}
See Appendix~\ref{poa_theorem_appendix}.
\end{IEEEproof}

\subsection{Efficiency Bound for Example Link-Cost Functions} 
\label{polynomial_subsection}
\textbf{Polynomial link-cost functions:}
We apply the above theorem to derive explicit expressions for the lower bound on the efficiency when the link-cost function is polynomial, i.e., $V(x)=ax^n$ where $a>0$. We start with the case of quadratic link-cost $V(x)=ax^2$. We then have $v(x)=2ax$ so that $v^{-1}(y)=\frac{y}{2a}$. Thus, using  (\ref{poa_bound_equn}), we obtain
\begin{eqnarray*}
\mathcal{E}(\{U_m\};V)
&\ge& \inf_{c>0} \frac{c\frac{c}{4a} - V(\frac{c}{4a})}{c \frac{c}{2a} - V(\frac{c}{2a})} \\
&=& \inf_{c>0} \frac{c\frac{c}{4a} - a(\frac{c}{4a})^2}{c \frac{c}{2a} - a(\frac{c}{2a})^2}  \\
&=& \inf_{c>0} \frac{\frac{c^2}{4a}(1 - \frac{1}{4})}{ \frac{c^2}{2a}(1 - \frac{1}{2})} \\
&=& \frac{3}{4}.
\end{eqnarray*}
Thus, when the link-cost is quadratic, the worst-case efficiency loss for any linear user pay-off is no more that $25\%$.

Similarly, suppose $V(x)=ax^3$. Then, using the bound (\ref{poa_bound_equn}) and a similar calculation, we obtain
\begin{eqnarray*}
\mathcal{E}(\{U_m\};V)\ \ge\ \frac{5}{4\sqrt{2}}\ \ge\ 0.88.
\end{eqnarray*}
Thus, the worst-case efficiency loss improves to  $12\%$ when the link-cost is cubic. In general, suppose the link-cost is polynomial of degree $n\ge2$, i.e., $V(x)=ax^n$, then the bound on efficiency is given by
\begin{eqnarray}
\label{poly_poa_equn}
\mathcal{E}(\{U_m\};V) &\ge& \left(\frac{1}{2}\right)^{\frac{n}{n-1}}\frac{2n-1}{n-1}.
\end{eqnarray}
The aforementioned lower bound is increasing as a function of $n$ and converges to $1$ as $n\rightarrow\infty$. Thus, if the link-cost can be modeled as $ax^n$, the efficiency loss reduces as $n$ increases.

\textbf{Exponential link-cost functions:} We next consider exponential link-cost functions of the form $V(x)=e^{ax}-(ax+1)$ where $a\ge1$. For these functions we have $v(x)=a(e^{ax}-1)$ so that $v^{-1}(y)= \frac{1}{a}\ln\left(\frac{y}{a}+1\right)$. Applying Theorem~\ref{poa_theorem} and simplifying, we obtain the following bound on efficiency:
\begin{eqnarray*}
\mathcal{E}(\{U_m\};V)
\ge \inf_{c>0} \frac{\left(\frac{c}{a}+1\right)\ln\left(\frac{c}{2a}+1\right) - \left(\frac{c}{2a}+1\right)^{\frac{1}{a}} + 1}{\left(\frac{c}{a}+1\right)\ln\left(\frac{c}{a}+1\right) - \left(\frac{c}{a}+1\right)^{\frac{1}{a}} + 1}.
\end{eqnarray*}
We numerically solve the above expression and (interestingly) find that
\begin{eqnarray}
\label{exponential_bound_equn}
\mathcal{E}(\{U_m\};V) &\ge&
\left\{
\begin{array}{ll}
{3}/{4} & \mbox{ for } a=1 \\
{1}/{2} & \mbox{ for } a>1.
\end{array}
\right.
\end{eqnarray}
Thus, exponential link-cost functions also yield favorable bound on efficiency. 

In summary, the above observations provide strong support for our proposed PAM-S mechanism when compared with the PAM-N mechanism of Section~\ref{price_anticipating_section} where the efficiency loss (for any $\{U_m\}$ and any $V$) is always $100\%$. 

\subsection{Worst-Case Bound on Efficiency} 
\label{worst_case_bound}
Although the class of polynomial link-cost functions yield favorable lower bounds on efficiency, we now show that there exists a family of link-cost functions $V_n$, $n\ge1$, such that the corresponding sequence of efficiency-bound converges to $0$ as $n\rightarrow\infty$. Thus, the {\em worst-case} efficiency bound, over all possible linear $\{U_m\}$ and over {\em all possible} $V$, is $0$.

To see this, let us first rewrite (\ref{poa_bound_equn}) by expressing $V$ in the integral form  $V(x)=\bigintssss_0^{x}v(\tau)d\tau$ to get
\begin{eqnarray*}
\mathcal{E}(\{U_m\};V) &\ge& \inf_{c>0} \frac{c v^{-1}(\frac{c}{2})- \bigintsss_{0}^{v^{-1}(\frac{c}{2})}v(\tau)d\tau   }{c v^{-1}(c)-\bigintsss_0^{v^{-1}(c)}v(\tau)d\tau} \\
&=:&\inf_{c>0} H(c,v).
\end{eqnarray*}
For a given $c$ and a marginal cost function for the link-supplier $v(\cdot)$, $H(c,v)$  can be geometrically interpreted with the aid of the illustration in Fig.\ref{geo_int_figure} as follows: the numerator in the formula for efficiency is the area of the region $\mathcal{A}_1$ (light shaded region) while the denominator is total area of $\mathcal{A}_1$ and $\mathcal{A}_2$ (shaded dark). We then have
\begin{eqnarray*}
H(c,v) = \frac{A_1}{A_1+A_2} = \frac{{A_1}/{A_2}}{1+{A_1}/{A_2}}
\end{eqnarray*}
where $A_i$ denotes the area of region $\mathcal{A}_i$ ($i=1,2$). In Fig.~\ref{geo_int_figure} we have used $e_1$ to denote $v^{-1}(\frac{c}{2})$; also, $f_1=v^{-1}(d_1)$ where $d_1$ is arbitrarily chosen in $\left(\frac{c}{2},c\right)$. Since $V$ is strictly convex and increasing, it follows that $v$ is strictly increasing.

Now, it is possible to construct a sequence of $v(\cdot)$ functions, say $\{v_n\}$, such that $e_n\downarrow0$, $d_n\downarrow \frac{c}{2}$ while $f_n\uparrow v^{-1}(c)$; an illustration of such a construction is depicted in Fig.~\ref{different_plots_figure}. Observe that along such a sequence we have $A_1\downarrow 0$ and $A_2\uparrow \frac{c}{2} v^{-1}(c)>0$. As a result we have $H(c,v_n)\rightarrow 0$ as $n\rightarrow\infty$. Thus, for any given $c$ it is possible to produce pathological link-cost functions whose efficiency-bounds are arbitrarily close to $0$. Therefore, it is not possible to guarantee a less-than-$100\%$ efficiency loss (i.e., a positive efficiency) when the class of all possible link-cost functions are considered. Nevertheless, bounding the efficiency for a fixed link-cost function is reassuring.

\begin{figure*}[t]
 \vspace{-4mm}
\centering
\subfigure[]{
\includegraphics[scale=0.32]{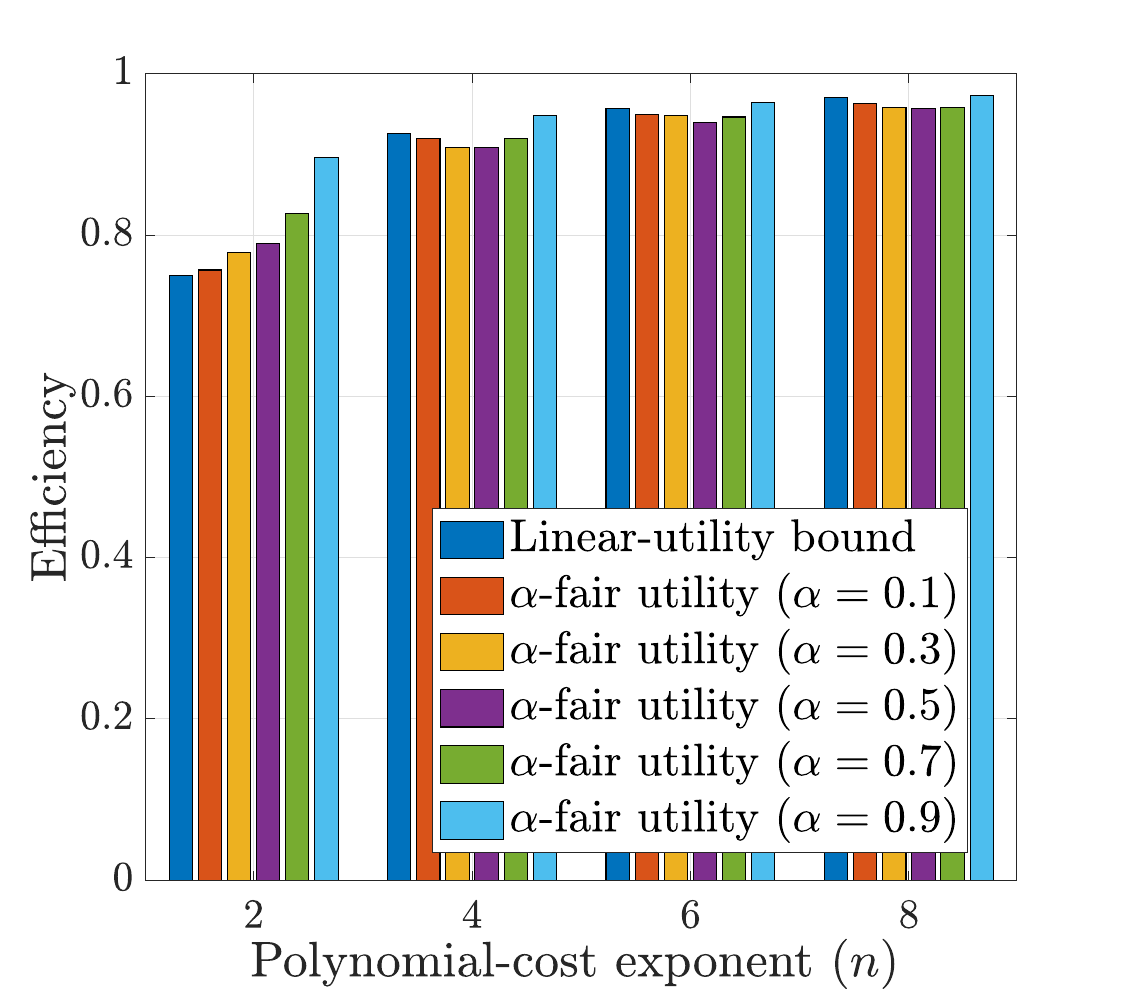}
        \label{efficiency_vs_n_for_diff_alpha_figure}
}
\hspace{-8mm}
\subfigure[]{
\includegraphics[scale=0.32]{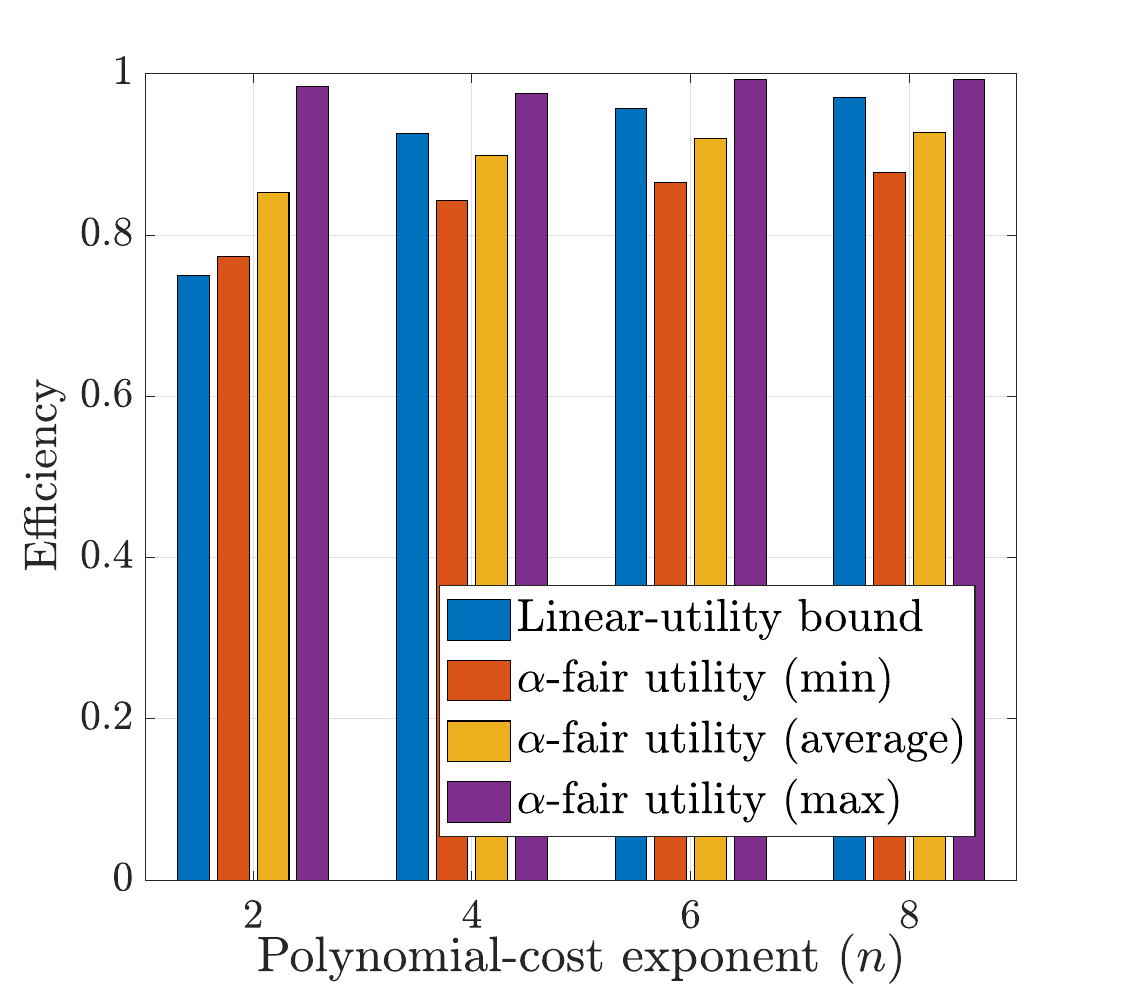}
        \label{efficiency_vs_n_random_alpha_figure}
}
\hspace{-8mm}
\subfigure[]{
\includegraphics[scale=0.32]{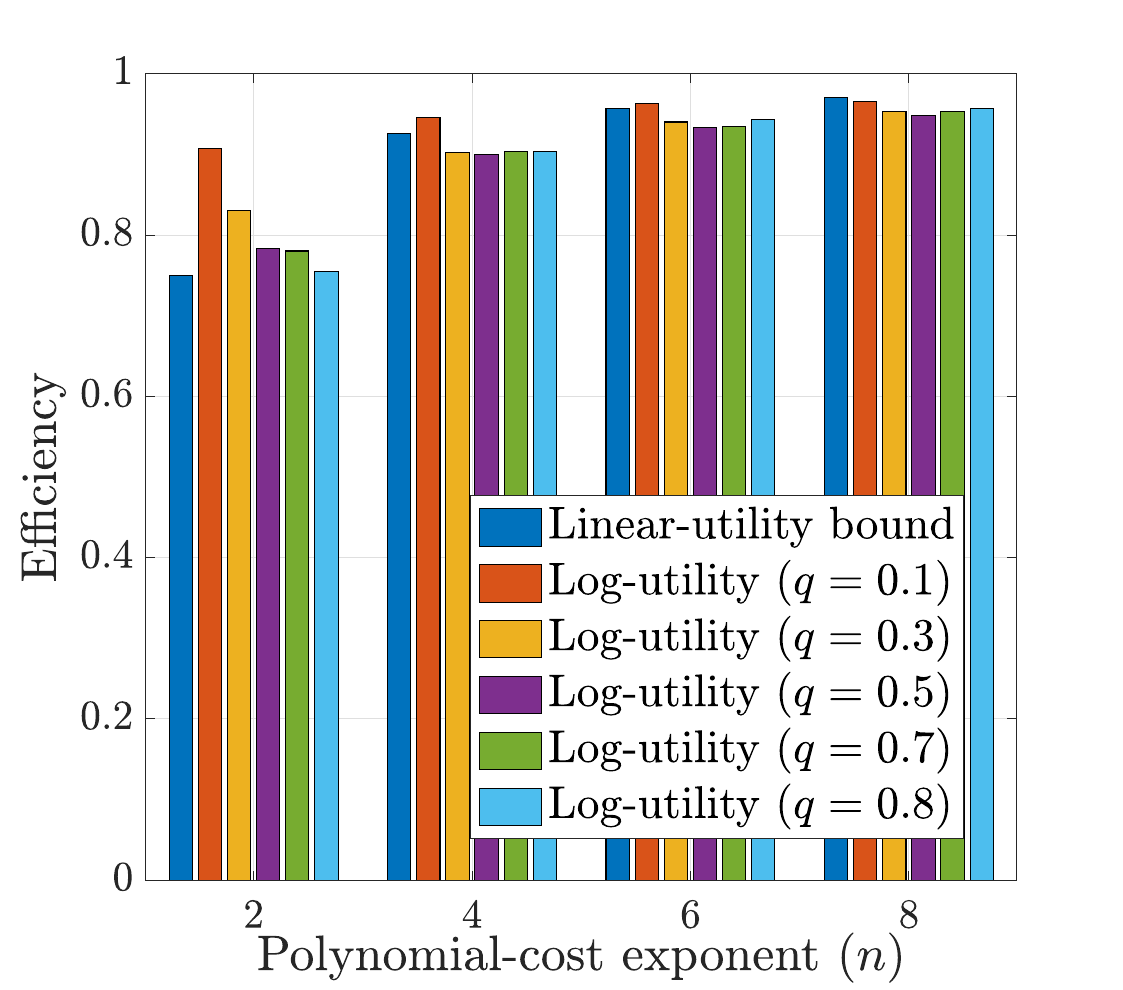}
        \label{efficiency_vs_q_bar_figure}
}
\subfigure[]{
\includegraphics[scale=0.32]{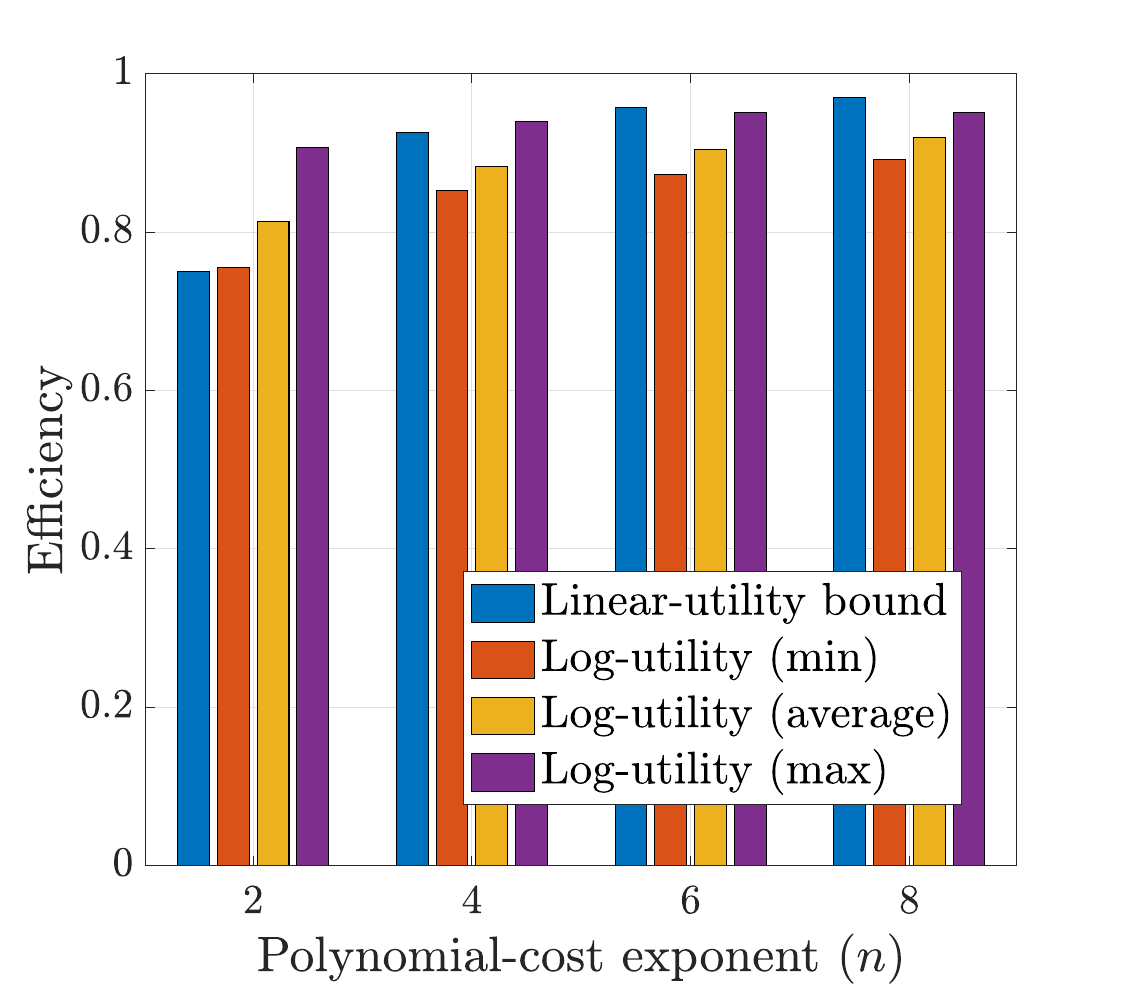}
        \label{efficiency_vs_q_random_figure}
}
\hspace{-8mm}
\subfigure[]{
\includegraphics[scale=0.32]{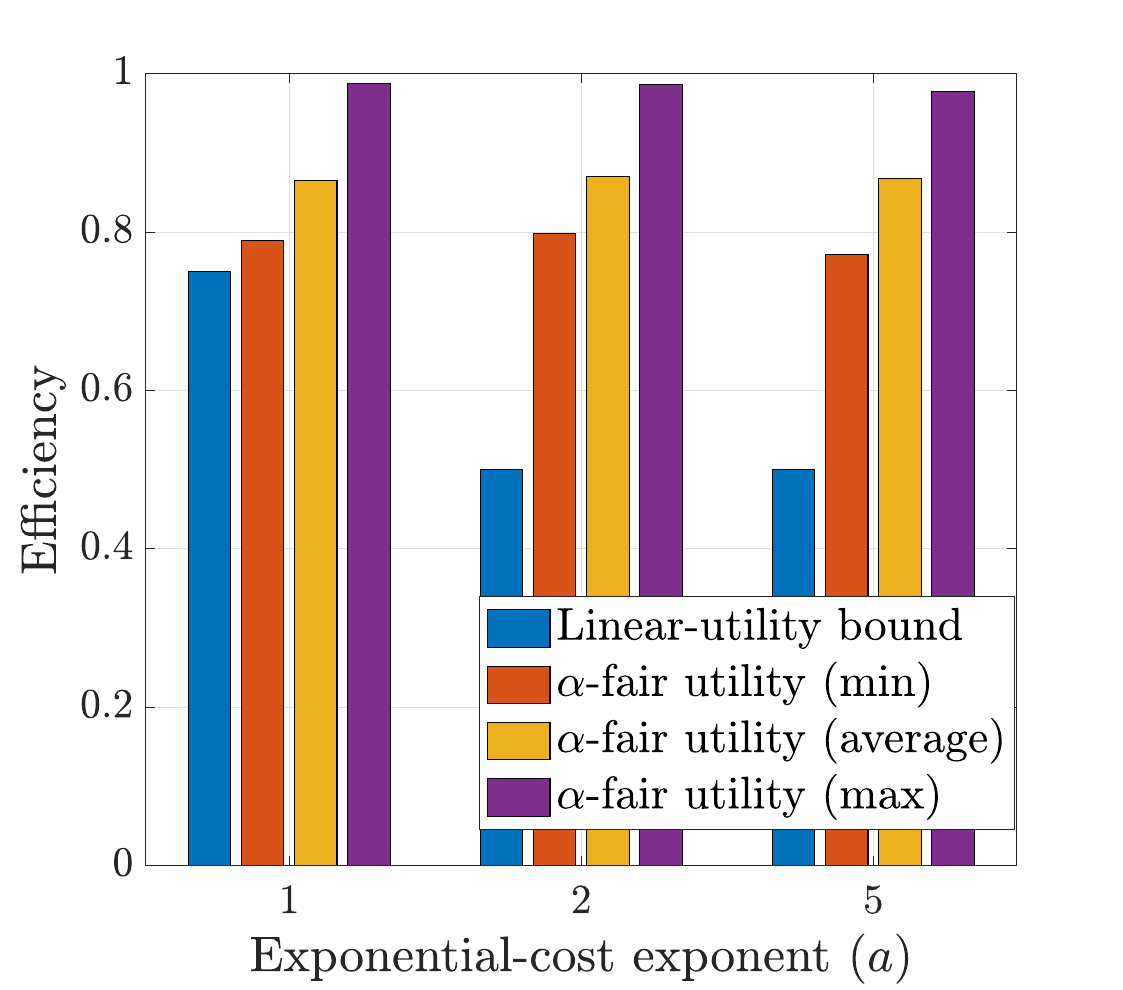}
        \label{expcost_efficiency_vs_a_figure}
}
\hspace{-8mm}
\subfigure[]{
\includegraphics[scale=0.32]{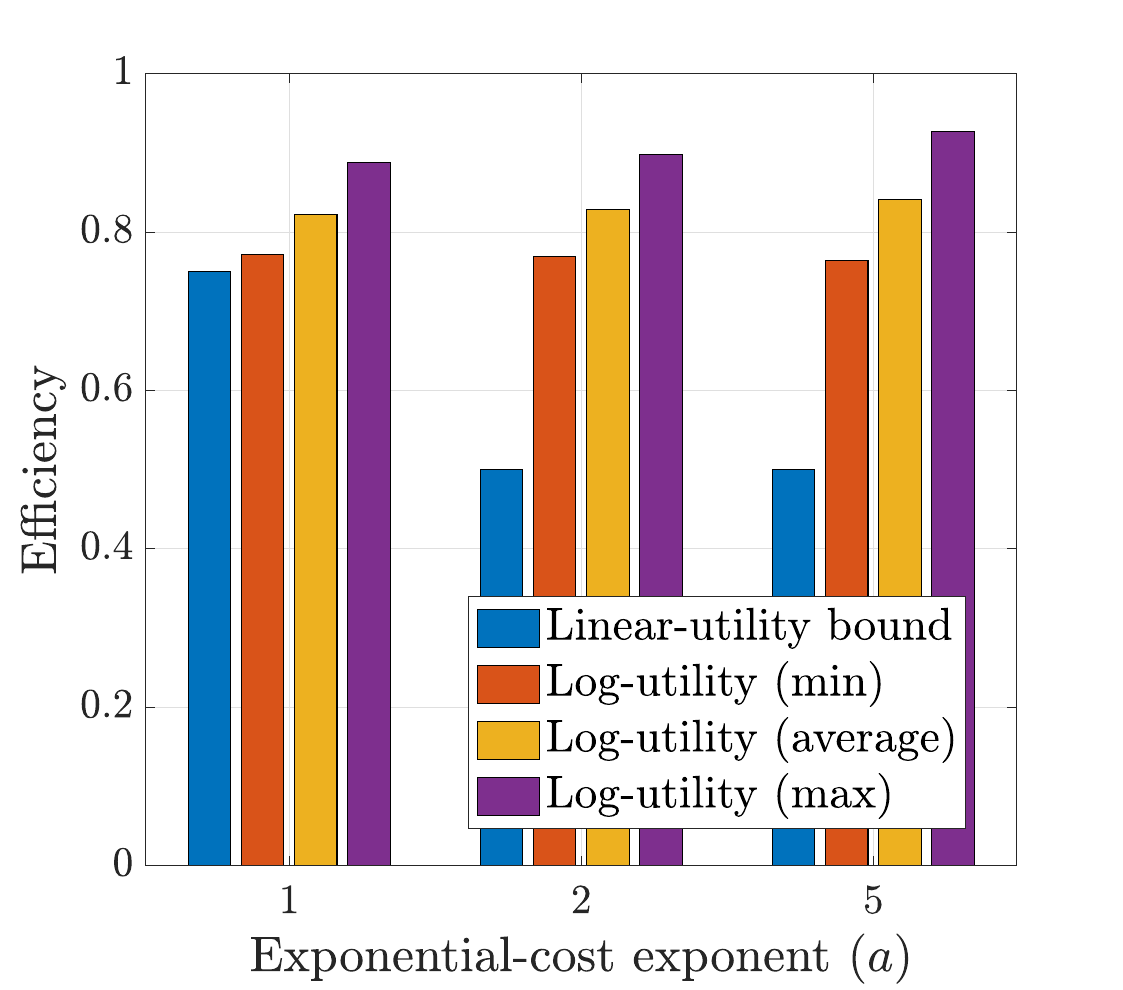}
       \label{expcost_efficiency_vs_a_random_figure}
}
\caption{\label{efficiency_figures} Efficiency of the proposed Stackelberg-game based mechanism (of Section~\ref{price_anticipation_delay_section}), evaluated for different combinations of user-utilities and link-cost functions. \subref{efficiency_vs_n_for_diff_alpha_figure} and \subref{efficiency_vs_n_random_alpha_figure}:  polynomial link-cost and $\alpha$-fair user utilities; \subref{efficiency_vs_q_bar_figure} and \subref{efficiency_vs_q_random_figure}: polynomial link-cost and $\log$ user utilities;   \subref{expcost_efficiency_vs_a_figure} exponential link-cost and $\alpha$-fair user utilities; and  \subref{expcost_efficiency_vs_a_random_figure} exponential link-cost and $\log$ user utilities.}
\vspace{-6mm}
\end{figure*}

\vspace{-2mm}
\section{Numerical Results}
\label{numerical_work_section}
Our theoretical result on efficiency of the proposed Stackelberg-game based mechanism presented in Theorem~\ref{poa_theorem} is limited to linear utility functions. In this section we conduct a detailed numerical study to characterize  efficiency for general utility functions. We specifically focus on the class of $\alpha$-fair and $\log$ utility functions described earlier (recall the remark following Theorem~\ref{stackelberg-existence-lemma}). For link-costs we use the polynomial and the exponential link-cost functions of Section~\ref{polynomial_subsection} so that the obtained results can be compared with the respective linear-utility bounds.

We first consider the polynomial link-cost function $V(x)=ax^n$. Since the corresponding linear-utility bound in (\ref{poly_poa_equn}) is not dependent on $a$, we simply fix $a=1$. We assume there are $M=5$ users in the system. We use $\alpha$-fair utility functions $U_m(x)=x^{1-\alpha_m}/{(1-\alpha_m)}$  to model user utilities (where $\alpha_m\in(0,1)$ for $m=1,2,\cdots,M$). We consider two approaches for studying the effect of $\{\alpha_m\}$ on efficiency. In the first approach we fix $\alpha_m=\alpha$ for all $m$ (\emph{identical user-utility scenario}), and compute efficiency by varying $\alpha$. In the second approach, $\alpha_m$ are assigned values in an i.i.d. fashion randomly from the interval $(0,1)$ (\emph{random user-utility scenario}). In the latter case we are interested in evaluating the min, max and the average values of the achieved efficiency.

In Fig.~\ref{efficiency_vs_n_for_diff_alpha_figure} we show the results obtained for the identical user-utility scenario.  Exponent $n$ of the polynomial link-cost function is varied along the $x$-axis while the achieved efficiency is depicted along the $y$-axis. For each $n$, we vary $\alpha$ from $0.1$ to $0.9$ in steps of $0.2$. For the purpose of comparison, in Fig.~\ref{efficiency_vs_n_for_diff_alpha_figure} we also present the linear-utility bound computed in (\ref{poly_poa_equn}). For $n=2$ the linear-utility bound of $3/4$ seems to apply to the case of $\alpha$-fair utilities as well for every value of $\alpha$. For other values of $n$, the bound appears to be applicable when $\alpha$ is larger. Although  smaller values of $\alpha$ yield a lower efficiency,  it is interesting to note that the efficiency values remain very close to the value of the bound derived for the linear utility functions.

In practice the user utilities need not be identical. We model this scenario by assigning values to $\alpha_m$ randomly from the interval $(0,1)$ in an i.i.d fashion. For a given assignment of $\{\alpha_m\}$ we compute the achieved efficiency which can be regarded as a random variable. We repeat the experiment for $100$ iterations of random assignment, computing efficiency in each iteration. In Fig.~\ref{efficiency_vs_n_random_alpha_figure} we depict the statistics of the obtained results in terms of the minimum, maximum, and the average efficiency achieved by the Stackelberg mechanism. Of particular interest is the max-statistics whose efficiency value is close to $1$, implying that for some configuration of $\{\alpha_m\}$ our Stackelberg-based rate-allocation yields a sum-utility which is as good as that achieved by the optimal allocation.

We continue our experiments with the polynomial link-cost function by replacing the $\alpha$-fair utility function with the $\log$-utility function $U_m(x)=\log(1+x^{q_m})$ where $q_m\in(0,1)$. The corresponding results are reported in Fig.~\ref{efficiency_vs_q_bar_figure} and \ref{efficiency_vs_q_random_figure}. Fig.~\ref{efficiency_vs_q_bar_figure} corresponds to the identical user-utility scenario where we fix $q_m=q$ for all $m$. The results are similar to  that in Fig.~\ref{efficiency_vs_n_for_diff_alpha_figure} except that here the efficiency decreases with the exponent $q$. Results of the random user-utility scenario, where $q_m$ are drawn i.i.d. randomly from $(0,1)$, are depicted in Fig.~\ref{efficiency_vs_q_random_figure}. Although the results are comparable with the case of $\alpha$-fair utility  (in Fig.~\ref{efficiency_vs_n_random_alpha_figure}), we observe that the max-statistics is not as promising as the results in  Fig.~\ref{efficiency_vs_n_random_alpha_figure}.

We finally conduct experiments with the exponential link-cost functions $V(x) = e^{ax}-(ax+1)$. For brevity, we present results only for the random user-utility scenario (however we note that the results of the identical user-utility case are similar to that obtained with polynomial link-cost function). In Fig.~\ref{expcost_efficiency_vs_a_figure} and \ref{expcost_efficiency_vs_a_random_figure} we present the statistics of efficiency (as a function of the exponent $a$) for $\alpha$-fair utility and log-utility, respectively. Also shown in these plots are the corresponding linear-utility bound on efficiency from (\ref{exponential_bound_equn}). From both plots we observe that the efficiency of the proposed mechanism is insensitive to the parameter $a$. Further, the linear-utility bound appears to hold for the class of $\alpha$-fair and $\log$ utility functions. As in the case of polynomial link-cost function, comparing the results in Fig.~\ref{expcost_efficiency_vs_a_figure} and \ref{expcost_efficiency_vs_a_random_figure} we find that the max-statistics is more favorable when the utility functions are $\alpha$-fair.

Finally, we would like a make a note regarding fairness of allocation, since from (\ref{stackelberg_unfair_allocation_equn}) we know that  linear utilities yield unfair allocation in the sense that, at equilibrium, except for one all users are allocated zero rate. Unlike the case with linear utilities, from our numerical work we find that the class of  $\alpha$-fair and $\log$ utility functions achieve a fairer rate allocation whereby non-zero rate is allocated to all users at equilibrium.

To summarize, we find that the proposed Stackelberg-game based resource-allocation mechanism performs consistently across the range of parameterized link-cost functions (polynomial and exponentials) and user utilities ($\alpha$-fair and $\log$) considered in our numerical study. In some cases the linear-utility bound is respected (e.g., $n=2$ in Fig.~\ref{efficiency_vs_n_for_diff_alpha_figure} to \ref{efficiency_vs_q_random_figure}, and for all values of $a$ in Fig~\ref{expcost_efficiency_vs_a_figure} and \ref{expcost_efficiency_vs_a_random_figure}). In scenarios where the linear-utility bound is not met, we observe that the values of the achieved efficiency are close to the respective  bounds. The linear-utility bound can thus serve as a good estimate for the system designer to take into consideration while designing a resource allocation network. Finally, summarizing results across all the plots in Fig.~\ref{efficiency_figures} we conclude that the efficiency of the proposed mechanism is at least $3/4$ ($75\%$) under any (link-cost and user-utility) scenario considered in our study.

\section{Conclusion}
\label{conclusion_section}
This paper was about double auction mechanisms and a proposal for a structured interaction to increase efficiency in the presence of strategic agents. The mechanism has application in data off-loading and network slicing markets. Data offloading is a good low-cost strategy that leverages existing auxiliary technology for handling the growth of mobile data. Technologies to enable such offloading are now available \cite{european_study,cisco}. Network slicing is expected to open up new business opportunities for mobile operators who can slice their physical resources and lease them to tenants or virtual network operators. Since in both examples the resulting markets are resource trading markets, suitable compensation mechanisms have to be put in place to encourage trading of the physical resources. It is natural that the agents involved are strategic. This paper demonstrates that mechanisms for trading resources should be designed with some care. An earlier work proposed a data offloading mechanism (collect bids, allocate offloading amounts, and distribute payments) and designed an iterative procedure to get the system to a competitive equilibrium where all agents benefited, if all agents were price-taking. We showed that if the agents are price-anticipating, this benefit completely disappears and the efficiency loss is 100\%. New mechanisms are thus needed when all agents are price-anticipating. We proposed a simple Stackelberg formulation with the supplying agent as a lead player. The resulting mechanism structures the interactions and alleviates the problem to some extent. The efficiency is lower bounded in terms of the true link cost function. The efficiency loss is 25\% for quadratic link costs (efficiency = 0.75). While there are link cost functions for which the efficiency loss, even in the Stackelberg formulation, is close to 100\%, these appear to be pathological cases. The proposed mechanism with link suppliers as lead players will likely have tolerable efficiency loss for most real link cost functions and arbitrary but linear user pay-offs. This is to be contrasted with 100\% efficiency loss for the price-anticipating mechanism. Going beyond our scalar bid per resource, our proposal also suggests an interesting open problem for implementation theorists. Does the minimum signaling dimension for social welfare maximization (in the Stackelberg equilibrium solution concept) strictly decrease?

\bibliographystyle{IEEEtran}
\bibliography{IEEEabrv,network-utility}

\vspace{-5mm}
\appendices
\label{proofs_section}
\label{app:price-anticipating-proofs}

\section{Proof of Lemma~\ref{lagrange_multipliers_lemma}}
\label{lagrange_multipliers_lemma_appendix}
The Lagrangian for the problem NETWORK in (\ref{network_equn}) is given by
\begin{eqnarray*}
L(\mathbf{x},\mathbf{y},\lambda,\bm{\mu})
&=& \vc{\sum}{}{m}
p_m \log(x_m) - \vc{\sum}{}{m}\frac{ y_m^2}{2\beta_m} \\
&& \hspace{2mm}-\lambda\Big(\vc{\sum}{}{m}y_m-C\Big) - \underset{m}{\sum}\mu_m\Big(x_m-y_m\Big)
\end{eqnarray*}
where $\lambda$ and $\bm{\mu}=(\mu_1,\cdots,\mu_m)$ are the Lagrange multipliers associated with the constraints in (\ref{net1_capacity_constraint_equn}) and (\ref{net1_matching_constraint_equn}), respectively. The optimality (KKT) conditions include
\begin{eqnarray}
\label{dl_dxm_equn}
\frac{\partial L}{\partial x_m} &=& \frac{p_m}{x_m}-\mu_m=0 \ \forall m \\
\label{dl_dym_equn}
\frac{\partial L}{\partial y_m} &=& - \frac{y_m}{\beta_m} -\lambda + \mu_m =0 \ \forall m
\end{eqnarray}
along with primal feasibility ((\ref{net1_capacity_constraint_equn}), (\ref{net1_matching_constraint_equn}) and (\ref{net1_nonnegative_constraint_equn})), dual feasiblity ($\lambda\ge0$ and  $\bm{\mu}\ge \textbf{0}$), and the following complementary slackness conditions:
\begin{eqnarray}
\label{capacity_slackness_equn}
\lambda\Big(\vc{\sum}{}{m}y_m-C\Big) &=& 0 \\
\label{mu_slackness_equn}
\mu_m\Big(x_m-y_m\Big) &=& 0\ \forall m.
\end{eqnarray}

Note that the objective in (\ref{net1_objective_equn}) is strictly increasing with $x_m$; hence, at optimality it should be that $x_m=y_m$ for all $m$. As in the proof of Theorem~\ref{competitive_equilibrium_theorem}  (see \emph{Step-4}), this observation can also be argued formally using the KKT conditions. Suppose at optimality we have  $x_m<y_m$ for some $m$. Then, the slackness condition in (\ref{mu_slackness_equn}) implies $\mu_m=0$, using which in (\ref{dl_dym_equn}) we obtain $y_m={-\beta_m\lambda}$. However, non-negativity constraint on all the variables forces $\beta_m\lambda=0$, yielding $0\le x_m = y_m=0$, which is a contradiction. Thus, at optimality we must have $x_m=y_m\ \forall m$. Now, using  (\ref{dl_dxm_equn}) and (\ref{dl_dym_equn}) in the above expression we have
\begin{eqnarray}
\label{xy_equality_equn}
x_m=\frac{p_m}{\mu_m} = \beta_m\Big(\mu_m-\lambda\Big) = y_m\ \forall m.
\end{eqnarray}
Solving this for $\mu_m$ we obtain (as required; recall (\ref{mu_equn}))
\begin{eqnarray}
\mu_m = \frac{\lambda+\sqrt{\lambda^2+4\frac{p_m}{\beta_m}}}{2}\ \forall m.
\end{eqnarray}

\begin{figure}[t]
\vspace{-5mm}
 \centering
\includegraphics[scale=0.3]{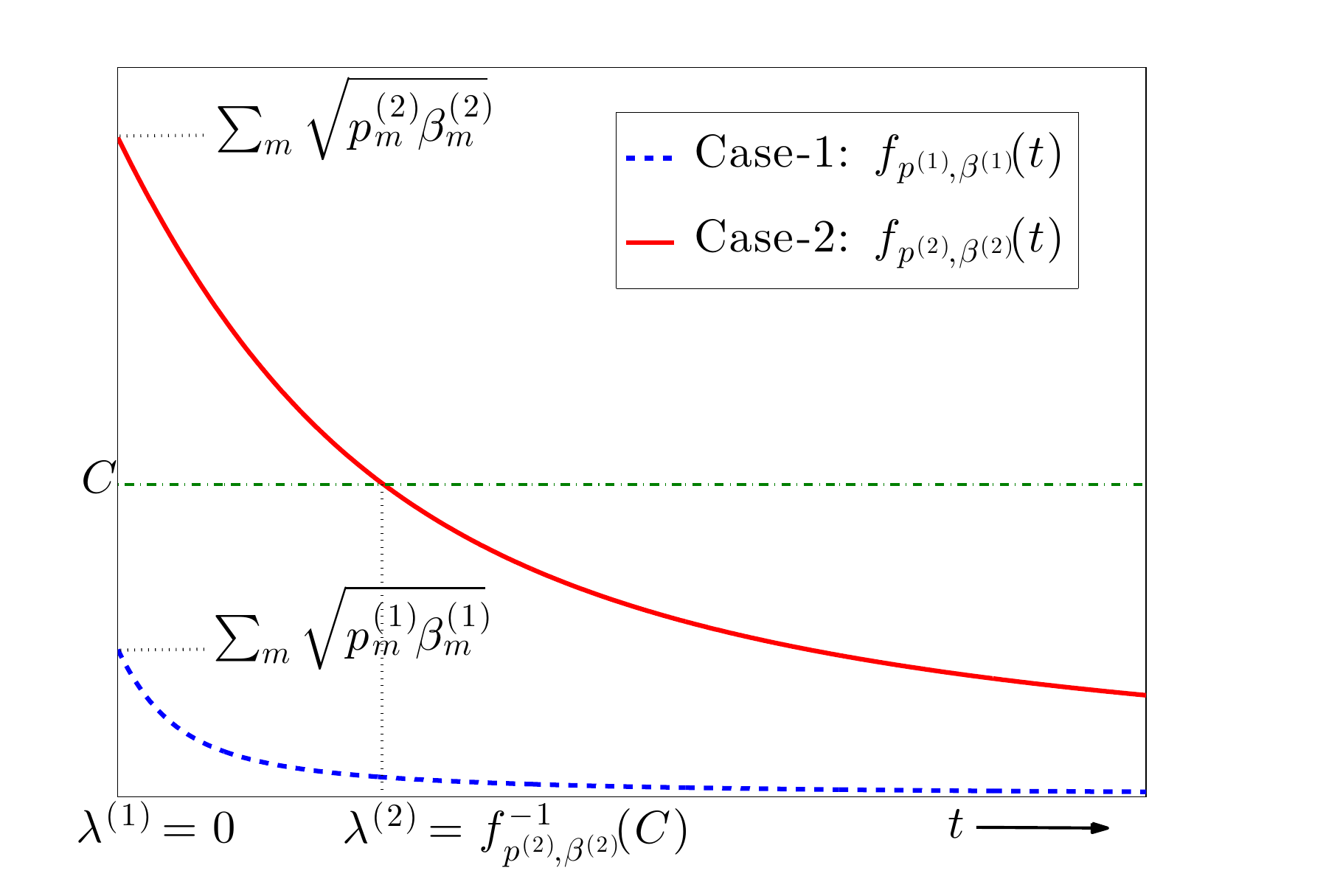}
 \caption{Illustration of the two cases that are possible depending on whether $f_{\textbf{p},\bm{\beta}}(0)=\vc{\sum}{}{m}\sqrt{{p_m}{\beta_m}}\le C$ (\emph{Case-1}) or otherwise (\emph{Case-2}). In the above depiction $(\textbf{p}^{(1)},\bm{\beta}^{(1)})$ and  $(\textbf{p}^{(2)},\bm{\beta}^{(2)})$ are such that they satisfy \emph{case-1} and \emph{case-2}, respectively, so that $\lambda^{(1)}:=\lambda({\textbf{p}^{(1)},\bm{\beta}^{(1)}})=0$ and $\lambda^{(2)}:=\lambda({\textbf{p}^{(2)},\bm{\beta}^{(2)}})=f_{\textbf{p}^{(2)},\bm{\beta}^{(2)}}^{-1}(C)$.
 \label{lbd_illustration_fig}}
 \vspace{-4mm}
 \end{figure}

We next proceed to obtain $\lambda$. For this, using the above expression in (\ref{xy_equality_equn}) and summing over all $m$, we obtain
\begin{eqnarray}
\label{xfyc_equn}
\underset{m}{\sum} x_m = f_{\textbf{p},\bm{\beta}} (\lambda) = \underset{m}{\sum} y_m \le C,
\end{eqnarray}
where the function $f_{\textbf{p},\bm{\beta}}$ is as in (\ref{f_func_equn}), and the inequality is simply due to the capacity constraint in (\ref{net1_capacity_constraint_equn}). Note that, $f_{\textbf{p},\bm{\beta}}(\lambda)$ as a function of $\lambda$ is strictly decreasing with $f_{\textbf{p},\bm{\beta}}(0)= \vc{\sum}{}{i}\sqrt{{p_i}{\beta_i}}$ (see Fig.~\ref{lbd_illustration_fig} for an illustration). Also,  $\lim_{\lambda\rightarrow\infty}f_{\textbf{p},\bm{\beta}}(\lambda)= 0$. Two cases are possible (recall (\ref{lbd_equn})) as follows.

 \emph{Case-1} ($f_{\textbf{p},\bm{\beta}}(0)\le C$): In this case $\lambda=0$ alone satisfies (\ref{xfyc_equn}) while ensuring the slackness condition in (\ref{capacity_slackness_equn}).

\emph{Case-2} ($f_{\textbf{p},\bm{\beta}}(0)>C$): In this case we require $\lambda>0$, since $\lambda=0$ cannot satisfy (\ref{xfyc_equn}). However,  $\lambda>0$  immediately implies $\vc{\sum}{}{m}y_m=C$ (see (\ref{capacity_slackness_equn})). Hence, we set $\lambda=f^{-1}_{\textbf{p},\bm{\beta}}(C)$.

\noindent
See  Fig.~\ref{lbd_illustration_fig} for an illustration of the above two cases.
\QEDB

\section{Proof of Theorem~\ref{pa_NE_theorem}}
\label{pa_NE_theorem_appendix}
We will first show that $(\textbf{p}^o,\bm{\beta}^o)$ is a Nash equilibrium. For this, note that once the link-supplier fixes his bids to $\bm{\beta}^o=\textbf{0}$, then for any vector of user bids $\textbf{p}\ge\bm{0}$ the system operates in the  regime $\vc{\sum}{}{i}\sqrt{{p_i}{\beta_i^o}}\le C$. Thus, using the first expression in (\ref{user_payoff_equn}), for any $\overline{p}_m>0$, we have
\begin{eqnarray*}
Q_m(\overline{p}_m,\mathbf{p}^o_{-m},\bm{\beta}^o)
&=& U_m(0)-\overline{p}_m \\
&<& U_m(0) \\
&=& Q_m({p}_m^o,\mathbf{p}^o_{-m},\bm{\beta}^o).
\end{eqnarray*}
Thus, unilateral deviation from $p_m^o$ is not beneficial for user $m$ ($\forall m$). Similarly, for any $\overline{\bm{\beta}}$ such  that $\overline{\beta}_m>0$  for some $m$, we have
\begin{eqnarray*}
Q_L(\overline{\bm{\beta}},\mathbf{p}^o)
= -V(0) + \underset{m}{\sum} p^o_m
= Q_L(\bm{\beta}^o,\mathbf{p}^o)
\end{eqnarray*}
To obtain the above, note that since the users\textquotesingle\ payments are zero, from (\ref{link_payoff_equn}), the first expression applies. Any other value of $\beta_m$ does not strictly increase the pay-off of the link-supplier.
Thus, $(\textbf{p}^o,\bm{\beta}^o)$ is a Nash equilibrium.

We now prove the uniqueness of the Nash equilibrium. Let $(\textbf{p}^*,\bm{\beta}^*)$ be a Nash equilibrium. Suppose $p_m^*>0$ for some $m$. Then, if $\beta_m^*=0$ (recalling (\ref{user_payoff_equn})) the pay-off to user $m$ is
\begin{eqnarray*}
Q_m(p_m^*,\mathbf{p_{-m}^*},\bm{\beta}^*) &=& U_m(0)-p_m^*
< U_m(0)  \\
 &=& Q_m(0,\mathbf{p_{-m}^*},\bm{\beta}^*)
\end{eqnarray*}
which contradicts the assumption that $(\textbf{p}^*,\bm{\beta}^*)$ is a Nash equilibrium. On the other hand, if $\beta_m^*>0$, then the link-supplier can benefit by deviating to the bid $\bm{\beta}^o$. This is because, the rate-cost incurred by deviating to $\bm{\beta}^o$ is always strictly lower (since he now provides zero bandwidth). Also, the payment $\sum_m p_m^*$ accrued under $\beta_m^*>0$ may be better if  the system was not already in the regime $\sum_i\sqrt{p_i^*\beta_i^*}\le C$; if already in that regime the payment remains unchanged. Formally,
\begin{eqnarray*}
Q_L(\bm{\beta}^*,\textbf{p}^*) < -V(0) + \underset{m}{\sum} p_m^*
 = Q_L(\bm{\beta}^o,\textbf{p}^*)
\end{eqnarray*}
which is again a contradiction. Thus, $p^*_m=0\ \forall m$, i.e., $\textbf{p}^*=\textbf{0}$.

Now, suppose $\beta_m^*>0$ for some $m$.  User $m$ can benefit by making a small payment $\overline{p}_{m}$. Indeed choose a $\overline{p}_m$ satisfying
\begin{eqnarray}
\label{pbar_equn}
0<\overline{p}_m\le \min\Big\{{C^2}/{\beta_m^*} ,q_m\Big\}
\end{eqnarray}
where $q_m$ is the maximizer of the function
\begin{eqnarray}
\label{h_func_equn}
h(p_m)
&=& U_m\left(\sqrt{{p_m}{\beta_m^*}}\right) - p_m
\end{eqnarray}
over $p_m\ge0$. Note that $h(p_m)$ is strictly concave in $p_m$. Hence, $q_m$ is the unique solution to the optimality condition
\begin{eqnarray*}
U_m'\left(\sqrt{{q_m}{\beta_m^*}}\right) = 2\sqrt{{q_m}/{\beta_m^*}}.
\end{eqnarray*}
Since $U'(\cdot)$ is strictly decreasing with  $U'(0)>0$, we have $q_m>0$, thus enabling us to choose a $\overline{p}_m$ satisfying (\ref{pbar_equn}).  The $\min$ term in (\ref{pbar_equn}) involving $C^2/\beta_m^*$  is required to ensure that the first expression of (\ref{user_payoff_equn}) is applicable. Thus, we have
\begin{eqnarray*}
Q_m(p_m^*,\mathbf{p_{-m}^*},\bm{\beta}^*)
&=& U_m(0)\  (\mbox{since } {p}_{m}^*=0) \\
&<& U_m\left(\sqrt{{\overline{p}_m}{\beta_m^*}}\right) - \overline{p}_m  \\
&=&Q_m(\overline{p}_m,\mathbf{p_{-m}^*},\bm{\beta}^*)
\end{eqnarray*}
where the inequality is because the function $h(\cdot)$, being strictly concave, is strictly increasing until $q_m$. The above contradiction implies that $\beta_m^*=0\ \forall m$, i.e., $\bm{\beta}^*=\textbf{0}$. Hence, $(\textbf{p}^o,\bm{\beta}^o)$ is the only Nash equilibrium.
\QEDB

\vspace{-4mm}
\section{Proof of Lemma~\ref{rbeta_lemma}}
\label{rbeta_lemma_appendix}
Since the objective function in (\ref{pmbeta_equn}) is continuously differentiable and strictly concave (both are easy to check), it suffices to show that $p_m^{\bm{\beta}}$ of (\ref{eqn:pmbeta}) solves the following optimality equation:
\begin{eqnarray*}
U_m'\left(\sqrt{{p_m}{\beta_m}}\right)\frac{\sqrt{\beta_m}}{2\sqrt{p_m}} -1=0.
\end{eqnarray*}
Indeed, with $p_m=p_m^{\bm{\beta}}$ of (\ref{eqn:pmbeta}) plugged into the above expression we have
\begin{eqnarray*}
U_m'(r_{\beta_m}) \frac{\beta_m}{2r_{\beta_m}}-1=0
\end{eqnarray*}
and so $r_{\beta_m}$ satisfies $U'_m(r_{\beta_m})={2r_{\beta_m}}/{\beta_m}$. The case when $\beta_m=0$ is straightforward.
\QEDB

\vspace{-3mm}
\section{Proof of Theorem~\ref{stackelberg-existence-lemma}}
\label{stackelberg-existence-lemma_appendix}
We first show that the objective function of the problem in (\ref{beta_star_equn}) is continuous in $\beta_m$ for each $m$. We next argue that it is sufficient to consider $\argmax$ in (\ref{beta_star_equn}) over a compact set of $\bm{\beta}$ values. Then the proof is completed by invoking Weierstrass theorem. The details are as follows.

{\em Proof of continuity}:
To show the continuity of the objective in (\ref{beta_star_equn}) it suffices to prove that $r_{\beta_m}$ is continuous in $\beta_m$ (for all $m$). For simplicity we omit the subscript $m$ hereafter.  Let us first prove right-continuity of $r_{\beta}$ at $\beta = 0$. Clearly, since $r_{\beta}$ is the solution to $U'_m(r) = 2r/\beta$, we see that $r_{\beta} \geq 0$ and, moreover, since $U'_m$ is strictly decreasing we have
$r_{\beta} = \frac{U'_m(r_{\beta})}{2} \beta \leq \frac{U'_m(0)}{2} \beta$
from which it follows that $r_0 = 0$ so that $r_{\beta}$ is continuous at $\beta = 0$. To prove continuity at any $\beta > 0$, we show that $r_\beta$ is Lipschitz continuous, i.e.,
 $|r_{\beta_1} - r_{\beta_2}| \leq \frac{U_m'(0)}{2} |\beta_1 - \beta_2|.$
Without loss of generality assume $\beta_2 > \beta_1$ so that $r_{\beta_2} > r_{\beta_1}$. Then we  have (again since $U_m'(\cdot)$ is decreasing)
\begin{eqnarray*}
  r_{\beta_2} - r_{\beta_1} & = & U'_m(r_{\beta_2})\beta_2/2 - U'_m(r_{\beta_1}) \beta_1/2 \\
  & \leq & U'_m(r_{\beta_2})\beta_2/2 - U'_m(r_{\beta_2}) \beta_1/2 \quad \\
  & = & (U'_m(r_{\beta_2})/2) (\beta_2 - \beta_1) \\
  & \leq & (U'_m(0)/2) (\beta_2 - \beta_1).
\end{eqnarray*}
This establishes Lipschitz continuity.

{\em Proof that it suffices to search for $\beta_m$ in a bounded set}:
First note that, using the definition of $r_{\beta_m}$ and $p^{{\bf \beta}}_m$, we may write the objective function as
\begin{eqnarray*}
 \lefteqn{-V\left(\sum_m \frac{p^{{\bm \beta}}_m}{r_{\beta_m}/\beta_m} \right) + \sum_m p^{\bm \beta}_m}\hspace{20mm} \\
  & = & -V\left(\sum_m \frac{p^{{\bm \beta}}_m}{U'_m(r_{\beta_m})/2} \right)+ \sum_m p^{{\bm \beta}}_m \\
  & \leq & -V\left(\sum_m \frac{p^{{\bm \beta}}_m}{U'_m(0)/2} \right)+ \sum_m p^{{\bm \beta}}_m
\end{eqnarray*}
since $V(\cdot)$ is strictly increasing. From the assumption that $V(x)/x \rightarrow \infty$ as $x \rightarrow \infty$, we see that the last term in the right-hand side above is less than 0 for all $(p^{\bf \beta}_m, 1 \leq m \leq M)$ with $\sum_m p^{\bf \beta}_m > P$ for some bounded $P$. Since each $p^{\bf \beta}_m \geq 0$, we trivially have that $0 \leq p^{\bf \beta}_m \leq P$.

From the formula for $r_{\beta_m}$ and $p^{{\bf \beta}}_m$, we have $p^{\bf \beta}_m = r_{\beta_m} U_m'(r_{\beta_m})/2$, and hence $0 \leq r_{\beta_m} U_m'(r_{\beta_m})/2 \leq P$. Under the assumption $r U'_m(r) \rightarrow \infty$ as $r \rightarrow \infty$, we must then have $0 \leq r_{\beta_m} \leq R$ for some bounded $R$, and since $U'_m$ is strictly decreasing and strictly positive, we must have $U'_m(r_{\beta_m}) \geq U'_m(R) > 0$. Using this, we then have
\[
  \beta_m = p^{\bf \beta}_m / (U'_m(r_{\beta_m})/2)^2 \leq 4P/(U'_m(R))^2 < \infty.
\]
This completes the proof.
\QEDB

\section{Proof of Lemma~\ref{beta_solution_lemma}}
\label{beta_solution_lemma_appendix}
For simplicity, let $f(\bfb)$ denote the objective function in (\ref{beta_star_equn_R1}). Then, for any $\bfb\ge\bm{0}$ we have
\begin{eqnarray*}
f(\bfb)
&=&  -V\left(\sum_m \frac{\beta_m c_m}{2}\right) +\sum_m \frac{\beta_m c_m^2}{4} \\
&\le& -V\left(\sum_m \frac{\beta_m c_m}{2}\right) +\sum_m \frac{\beta_m c_m c_1}{4} \\
&=& -V\left(\frac{\tilde{\beta}_1c_1}{2}\right) +\frac{\tilde{\beta}_1c_1^2}{4} =  f(\tilde{\bfb}) \\
\end{eqnarray*}
where $\tilde{\bfb} = (\tilde{\beta}_1,\tilde{\beta}_2,\cdots,\tilde{\beta}_M)$ is such that $\tilde{\beta}_1=\sum_m \frac{\beta_m c_m}{c_1} $ while $\tilde{\beta}_m=0$ for $m=2,3,\cdots,M$. Thus, for any $\bfb\ge\bm{0}$ there exists a bid $\tilde{\bfb}\ge\bm{0}$ (as constructed above) using which the link-supplier can accrue a higher pay-off.  As a consequence, the optimization framework in (\ref{beta_star_equn_R1}) can be reduced to the following equivalent problem:
\begin{eqnarray*}
 \underset{\bm{\beta}=(\beta_1,0,\cdots,0) \geq \bm{0}}{\max}\left\{  -V\left( \frac{\beta_1 c_1}{2}\right) + \frac{\beta_1 c_1^2}{4}
\right\}.
\end{eqnarray*}
Solving for $\beta_1$ in the above expression (by taking derivative and equating to $0$) yields the result in (\ref{beta_m_star_equn_R1}).
\QEDB

\section{Proof of Theorem~\ref{poa_theorem}}
\label{poa_theorem_appendix}

Consider linear pay-offs of the form $U_m(x_m)=c_mx_m$ where $c_m>0$ ($m=1,2,\cdots,M$). Without loss of generality, assume that $c_1=\max_m \{c_m\}$. Then, recalling (\ref{xm_eq_rm_equn}), the utility at Stackelberg equilibrium can be written as
\begin{eqnarray}
\label{stackelberg_utility}
\lefteqn{\mbox{Stackelberg utility} } \hspace{10mm}\nonumber\\
&=& \sum_m U_m\left(x_{m}^{\bfb^*}\right) - V\left(\sum_m x_{m}^{\bfb^*}\right) \nonumber \\
&=&  U_1\left(v^{-1}\left(\frac{c_1}{2}\right)\right) - V\left(v^{-1}\left(\frac{c_1}{2}\right)\right) \nonumber \\
&=& c_1 v^{-1}\left(\frac{c_1}{2}\right) - V\left(v^{-1}\left(\frac{c_1}{2}\right)\right).
\end{eqnarray}

Next, the social optimal utility is obtained by solving
\begin{eqnarray*}
\max_{\bf{x} \geq \bf{0}} \left\{ \underset{m}{\sum} U_m\left(x_{m}\right) -  V\left(\sum_m x_{m}\right)  \right\}.
\end{eqnarray*}
Substituting for the linear pay-off functions and rearranging, the above problem can be alternatively expressed as
\begin{eqnarray*}
\max_{\bf{x} \geq \bf{0}} \left\{\sum_m c_m x_{m} - V\left(\sum_m x_{m}\right)  \right\}.
\end{eqnarray*}
Thus, the optimal rate allocation $x_{m}^s$ is given by
\begin{eqnarray*}
x_{m}^{s} &=&
\left\{
\begin{array}{ll}
v^{-1}(c_1) & \mbox{ if } m=1 \\
0 & \mbox{ otherwise.}
\end{array}
\right.
\end{eqnarray*}
Thus, the social optimal utility is given by
\begin{eqnarray}
\label{social_utility}
\mbox{Social utility} =  c_1 v^{-1}(c_1)-V\left(v^{-1}(c_1)\right).
\end{eqnarray}

From (\ref{stackelberg_utility}) and (\ref{social_utility}) we have (replacing $c_1$ by $c$)
\begin{eqnarray*}
\mathcal{E}(\{U_m\};V) =   \frac{c v^{-1}(\frac{c}{2})-V(v^{-1}(\frac{c}{2}))}{c v^{-1}(c)-V(v^{-1}(c))}.\end{eqnarray*}
The worst case bound on efficiency can be obtained by taking infimum over all $c>0$.
\QEDB

\vspace{-10mm}
\begin{IEEEbiography}
[{\includegraphics[width=1in,height =1.25in,clip,keepaspectratio]{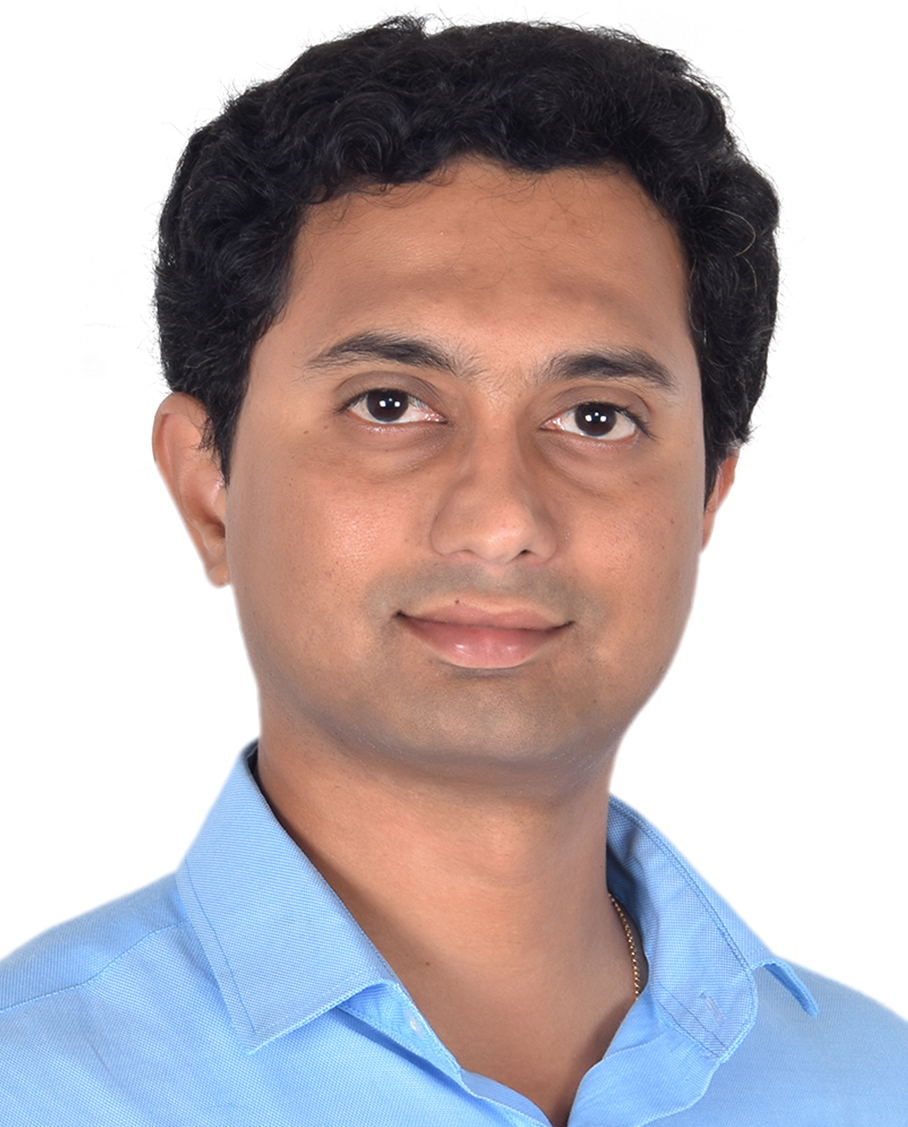}}]
{K.~P.~Naveen} (M'19) received the B.E.\ degree in ECE from the Visveswaraiah Technological University in 2005, and the Ph.D.\ degree from the Department of Electrical Communication Engineering, Indian Institute of Science (IISc), Bangalore in 2013. His work experience includes Sci/Engg. SC, ISRO Satellite Centre, Bangalore (2006 - 2007),  Post-doctoral Fellow, INRIA Saclay, France (2014 - 2015), and INSPIRE Faculty, Department of Electrical Engineering, Indian Institute of Technology Madras (2016 - 2017). Since July 2017 he is with the Department of Electrical Engineering, Indian Institute of Technology Tirupati. His research interests include modeling and performance analysis of wireless networks, network economics, game theory, and  optimal control.
\end{IEEEbiography}

\vspace{-5mm}
\begin{IEEEbiography}[{\includegraphics[width=1in,height =1.25in,clip,keepaspectratio]{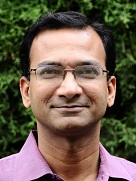}}]{Rajesh Sundaresan}
(S’96-M’00-SM’06) received the B.Tech. degree in electronics and communication from the Indian Institute of Technology Madras, India, and the M.A. and Ph.D. degrees in electrical engineering from Princeton University, Princeton, NJ, USA, in 1996 and 1999, respectively. From 1999 to 2005, he worked at Qualcomm Inc. on the design of communication algorithms for wireless modems. Since 2005, he has been with the Indian Institute of Science, Bangalore, India, where he is a Professor in the Department of Electrical Communication Engineering, an associate faculty in the Robert Bosch Centre for Cyber-Physical Systems, and a co-convenor of the Centre for Networked Intelligence. His interests include the areas of communication, computation, and control over networks. He was an Associate Editor for the IEEE Transactions on Information Theory for the period 2012-2015. 
\end{IEEEbiography}

\newpage

\section*{Supplementary Material}
\section{Proof of Theorem~\ref{competitive_equilibrium_theorem}}
\label{app:proof-of-CE}
The proof is based on Lagrangian technique. We now outline the key steps before going into the details. We first show that the  optimality conditions for SYSTEM  implies (C1)$-$(C3) that are required of a competitive equilibrium (CE); since the problem SYSTEM is convex, the existence result simply follows from the existence of a (primal and dual optimal) solution to the KKT conditions \cite{boyd_vandenberghe04convex_optimization,bertsekas_etal03convex_optimization}. For the second part, starting with (C1)$-$(C3) we deduce the optimality conditions for SYSTEM; the second part of the theorem then follows since KKT conditions in our present setting are both necessary and sufficient for optimality. Details follow.

\subsection{Proof of Existence of CE}

\noindent
\emph{\textbf{Step-1}: KKT conditions for SYSTEM}

The Lagrangian for SYSTEM in (\ref{system_equn}) is given by
\begin{eqnarray*}
L(\mathbf{x},\mathbf{y},\lambda,\bm{\mu})
&=&
\vc{\sum}{M}{m=1} U_m(x_m) - V\left(\vc{\sum}{M}{m=1} y_m\right)  \\
&& \hspace{2mm}-\lambda\Big(\vc{\sum}{}{m}y_m-C\Big) - \underset{m}{\sum}\mu_m\Big(x_m-y_m\Big)
\end{eqnarray*}
where, $\lambda$ and $\bm{\mu}=(\mu_1,\cdots,\mu_m)$ are the Lagrange multipliers associated with the constraints in (\ref{system_capacity_constraint_equn}) and (\ref{system_matching_constraint_equn}), respectively.
Defining $y=\underset{i}{\sum}y_i$, the optimality conditions are given by
\begin{eqnarray}
\label{um_optimality_equn}
\left.
\begin{array}{ll}
U_m'(x_m) = \mu_m & \mbox{ if } x_m>0 \\
U_m'(0) \le \mu_m & \mbox{ if } x_m=0
\end{array}
\right\}
\end{eqnarray}
\begin{eqnarray}
\label{vm_optimality_equn}
\left.
\begin{array}{ll}
V'\left(y\right) + \lambda = \mu_m  & \mbox{ if } y_m>0 \\
V'\left(y\right) + \lambda \ge \mu_m & \mbox{ if } y_m=0
\end{array}
\right\}
\end{eqnarray}
along with primal feasibility ((\ref{system_capacity_constraint_equn})$-$(\ref{system_nonnegative_constraint})), dual feasibility (i.e., $\lambda\ge0$, $\mu_m\ge0\ \forall m$) and complementary slackness conditions:
\begin{eqnarray}
\label{slackness1_equn}
\lambda\Big(\vc{\sum}{}{m}y_m-C\Big)
&=& 0 \\
\label{system_mu_slackness_equn}
\mu_m\Big(x_m-y_m\Big)
&=& 0\ \forall m.
\end{eqnarray}

\noindent
\emph{\textbf{Step-2}: Identifying a candidate competitive-equilibrium}

Since the problem is convex there exist primal and dual optimal points, $(\textbf{x}^s,\textbf{y}^s)$ and $(\lambda^s,\bm{\mu}^s)$, respectively, that together satisfy the above KKT conditions. Define $\textbf{p}^s$ and $\bm{\beta}^s$ as
\begin{eqnarray}
\label{p_define_equn}
p_m^s &=&
 x^s_m\mu_m^s  \\
\label{a_define_equn}
\beta_m^s &=&
\left\{
\begin{array}{ll}
\frac{y_m^s}{\left(\mu_m^s-\lambda^s\right)} &\mbox{ if } \mu_m^s\ne\lambda_s \\
0 & \mbox{ otherwise}.
\end{array}
\right.
\end{eqnarray}

\noindent
In the following
we will show that $\left(\textbf{p}^s,\bm{\beta}^s,\lambda^s,\bm{\mu}^s\right)$ is a competitive equilibrium, i.e., we verify (C1)$-$(C3).

\noindent
\emph{\textbf{Step-3}: Verifying (C1) and (C2)}

To verify (C1), we need to show that $p_m^s$ is optimal for the  problem of maximizing  $P_m(p_m;\mu_m^s)$ in (\ref{pt_user_payoff_equn}) over $p_m\ge0$, the optimality condition for which is given by
\begin{eqnarray*}
\begin{array}{ll}
U_m'\left(\frac{p_m}{\mu_m^s}\right) = \mu_m^s & \mbox{ if } p_m > 0 \\
U_m'(0) \le \mu_m^s & \mbox{ if } p_m=0.
\end{array}
\end{eqnarray*}
From (\ref{um_optimality_equn}) and (\ref{p_define_equn}) it follows that $p_m^s$ satisfies the above condition, thus implying (C1).

Similarly, the optimality condition for maximizing the  link pay-off function, $P_L\Big(\bm{\beta};(\bm{\mu}^s,\lambda^s)\Big)$ in (\ref{pt_link_payoff_equn}) over $\bm{\beta}\ge\textbf{0}$ is:

\begin{itemize}
\item if $\beta_m>0$
\begin{eqnarray}
\label{case1_equn}
V'\left(\vc{\sum}{}{i} \beta_i\Big({\mu_i^s-\lambda^s}\Big)\right) \Big(\mu_m^s-\lambda^s\Big) = \Big(\mu_m^s-\lambda^s\Big)^2;
\end{eqnarray}

\item if $\beta_m=0$
\begin{eqnarray}
\label{case2_equn}
V'\left(\vc{\sum}{}{i} \beta_i\Big({\mu_i^s-\lambda^s}\Big)\right) \Big(\mu_m^s-\lambda^s\Big) \ge \Big(\mu_m^s-\lambda^s\Big)^2.
 \end{eqnarray}
\end{itemize}

We show that $\beta_m^s$ in (\ref{a_define_equn}) satisfies the above. Suppose $\beta_m^s>0$ then it should be that  $\mu_m^s\ne\lambda^s$. Thus, with $\bm{\beta}=\bm{\beta}^s$, expression (\ref{case1_equn}) simplifies to
\begin{eqnarray}
\label{C2_verification_equn}
V'\left(\vc{\sum}{}{i} \beta_i^s\Big({\mu_i^s-\lambda^s}\Big)\right) + \lambda^s
= \mu_m^s=:\mu^s
\end{eqnarray}
which follows from the first part of  (\ref{vm_optimality_equn}). On the other hand, if $\beta_m^s=0$ then either $\mu_m^s=\lambda_s$ or $y_m^s=0$ (or both). If the former is true then (\ref{case2_equn}) holds trivially, while in the latter case (\ref{case2_equn}) follows from the second part of  (\ref{vm_optimality_equn}).  Thus, condition (C2) is verified.

\noindent
\emph{\textbf{Step-4}: Verifying (C3)}

To verify (C3-a) we need to prove that $\textbf{x}^s=\textbf{y}^s$. Evidently, since the objective in (\ref{system_objective_equn}) is strictly increasing in $x_m$ (owing to the strictly-increasing condition imposed on $U_m$), at optimality it should be that $\textbf{x}^s=\textbf{y}^s$. For completeness, we formally prove this result using the optimality conditions. Suppose $0\le x_m^s<y_m^s$ for some $m$. Then, the slackness condition in (\ref{system_mu_slackness_equn}) immediately implies that $\mu_m^s=0$, so that from (\ref{vm_optimality_equn}) we obtain $V'(y^s)+\lambda^s=0$. Since $V$ is strictly increasing and $y_m^s>0$, we have $V'(y^s)>0$ yielding $\lambda^s<0$ which is a contradiction. Thus, $x_m^s=y_m^s\ \forall m$.

Since (C2) is already verified, from (\ref{C2_verification_equn}) we see that  we have $\mu_m^s=\mu^s$ for all $m\in\mathcal{M}^s$, where  $\mathcal{M}^s=\{m:\mu_m^s\ne \lambda^s\}$. Thus, (summing the expression in (\ref{p_define_equn}) over all $m$) we have
\begin{eqnarray}
\label{sum_xm_equn}
\underset{m}{\sum}x_m^s = \underset{m}{\sum}\frac{p^s_m}{\mu_m^s}
&=&\underset{m\in\mathcal{M}^s}{\sum}\frac{p^s_m}{\mu^s} + \underset{m\notin\mathcal{M}^s}{\sum}\frac{p^s_m}{\mu_m^s} \nonumber\\
&=&\frac{1}{\mu^s} {\underset{m}{\sum} p^s_m}.
\end{eqnarray}
The last equality is obtained by noting that $p^s_m=0$ for $m\notin\mathcal{M}^s$. This is because, when $\mu_m^s=\lambda^s$, from (\ref{vm_optimality_equn}) we see that $y_m^s=0$ (as $y_m^s>0$ leads to the contradiction $V'(y^s)=0$); now, since $x_m^s=y_m^s$ (from (C3-a)) we obtain $p_m^s=0$  from (\ref{p_define_equn}). This establishes (\ref{sum_xm_equn}).

Similarly, using (\ref{a_define_equn}) and by splitting the following sum  over $m \in \mathcal{M}^s$ and $m \notin \mathcal{M}^s$ separately,  we obtain
\begin{eqnarray}
\label{sum_ym_equn}
\underset{m}{\sum} y_m^s
= \underset{m}{\sum}{\beta_m^s} {\Big(\mu_m^s-\lambda^s\Big)}
= \Big(\mu^s-\lambda^s\Big) \underset{m\in\mathcal{M}^s}{\sum} \beta_m^s.
\end{eqnarray}
Note that, unlike in (\ref{sum_xm_equn}), the  summation in the final term above is restricted to $m\in\mathcal{M}^s$. This is because, in this case whenever $\mu_m^s\ne\lambda^s$ it is not necessary that $\beta_m^s=0$. We need to consider two cases:
\begin{itemize}
\item Suppose $\lambda^s=0$. Then, since $\underset{m}{\sum}x_m=\underset{m}{\sum}y_m$ (recall (C3-a)), solving for $\mu^s$ from (\ref{sum_xm_equn}) and (\ref{sum_ym_equn}) we obtain
\begin{eqnarray*}
\mu^s = {\underset{m}{\sum}p_m^s}/{\widehat{C}^s}
\end{eqnarray*}
where $\widehat{C}^s$ is as in (\ref{C0_equn}) (but with $(\textbf{p},\bm{\beta})$ replaced with $(\textbf{p}^s,\bm{\beta}^s)$). Using the above in  (\ref{sum_xm_equn}) yields $\underset{m}{\sum}x_m^s = \widehat{C}$. Since $\underset{m}{\sum}x_m^s\le C$,  we have $\widehat{C}^s\le C$ under this case.

\item Suppose $\lambda^s>0$. Then the slackness condition in (\ref{slackness1_equn}) immediately implies $\underset{m}{\sum}y_m=C=\underset{m}{\sum}x_m$. Thus, from (\ref{sum_xm_equn}), we have $\mu^s = {\underset{m}{\sum}p_m^s}/{C}.$
Using this in (\ref{sum_ym_equn}) and simplifying for $\lambda^s$ we obtain
 \begin{eqnarray*}
 \lambda^s =\left(1-\left(\frac{C}{\widehat{C}^s}\right)^2\right) \frac{\vc{\sum}{}{i} p_i^s}{C}.
\end{eqnarray*}
Note that, since $\lambda^s>0$, we have $\widehat{C}^s>C$ in this case.
\end{itemize}
Results from the above two conditions can be compactly expressed as (\ref{mu1_equn}) and (\ref{lambda_equn}); thus (C3-b) and (C3-c) are verified.

\subsection{Proof of Optimality of CE}
Let $(\textbf{p}^c,\bm{\beta}^c,\lambda^c,\bm{\mu}^c)$ be a competitive equilibrium. Define rate-vectors $\textbf{x}^c$ and $\textbf{y}^c$ as $x_m^c=\frac{p_m^c}{\mu_m^c}$ and $y_m^c=\beta_m^c(\mu_m^c-\lambda^c)$. We will show that $(x^c,y^c,\lambda^c,\mu^c)$ satisfies the KKT conditions for the problem SYSTEM (recall \emph{Step-1}).

From (C1) it follows that ${p}_m^c$ is optimal for the user problem of maximizing the pay-off function $P_m(p_m;\mu_m^c)$ in (\ref{pt_user_payoff_equn}) over all $p_m\ge0$. Thus we have
\begin{eqnarray*}
\begin{array}{ll}
U_m'\left(\frac{p_m^c}{\mu_m^c}\right) = \mu_m^c & \mbox{ if } p_m^c > 0 \\
U_m'(0) \le \mu_m^c & \mbox{ if } p_m^c=0.
\end{array}
\end{eqnarray*}
With ${x}_m^c:=\frac{p_m^c}{\mu_m^c}$ the above expression is identical to the KKT condition in (\ref{um_optimality_equn}).

Similarly, (C2) implies that $\bm{\beta}^c$ is optimal for the link's problem of maximizing $P_L\Big(\bm{\beta};(\bm{\mu}^c,\lambda^c)\Big)$ in (\ref{pt_link_payoff_equn}) over $\bm{\beta}\ge\textbf{0}$. Thus, $\bm{\beta}^c$ satisfies the optimality conditions in (\ref{case1_equn}) and (\ref{case2_equn}), but with $(\lambda^s,\bm{\mu}^s)$ replaced by $(\lambda^c,\bm{\mu}^c)$.
\begin{itemize}
\item Suppose $y_m^c=\beta_{m}^c(\mu_m^c-\lambda^c)>0$ then $\beta_m^c>0$ and $\mu_m^c\ne\lambda^c$; the optimality condition in this case simplifies to the expression in (\ref{C2_verification_equn}) (but again with $(\bm{\beta}^s,\lambda^s,\bm{\mu}^s)$ replaced by $(\bm{\beta}^c,\lambda^c,\bm{\mu}^c)$).  Then, substituting $y_m^c=\beta_m^c(\mu_m^c-\lambda^c)$, we obtain the first part of the KKT condition in (\ref{vm_optimality_equn}) is satisfied.

\item Suppose $y_m^c=0$ then either $\mu_m^c=\lambda^c$ or $\beta_m^c=0$ (or both). $\mu_m^c=\lambda^c$ case trivially satisfies the optimality conditions in (\ref{case1_equn}) and (\ref{case2_equn}). If $\beta_m^c=0$ then the second part of (\ref{vm_optimality_equn}) follows from (\ref{case2_equn}).
\end{itemize}

For the slackness conditions in (\ref{slackness1_equn}) and (\ref{system_mu_slackness_equn}), note that condition (C3-1) in (\ref{C3_1_equn}) implies  $x_m^c=y_m^c$ ($\forall m$) so that (\ref{system_mu_slackness_equn}) holds immediately. To show (\ref{slackness1_equn}), first suppose that $\lambda^c>0$. Then, from (C3-c) in (\ref{lambda_equn}) we have $\widehat{C}^c>C$ (where $\widehat{C}^c$ is as in (\ref{C0_equn}) but with $(\textbf{p},\bm{\beta})$ replaced by $(\textbf{p}^c,\bm{\beta}^c)$). Using the above condition in (\ref{mu1_equn}) of (C3-b)  we obtain $\mu=\sum_{i} p_i^c /{C}$. Thus
\begin{eqnarray*}
\underset{m}{\sum} y_m^c
= \underset{m}{\sum} x_m^c
&=& \underset{m\in\mathcal{M}}{\sum} \frac{p_m^c}{\mu_m^c} + \underset{m\notin\mathcal{M}}{\sum} \frac{p_m^c}{\mu_m^c} \frac{\underset{m}{\sum} {p_m^c}}{\mu} = C,
\end{eqnarray*}
where we have used the fact that $p_m^c = 0$ for $m \notin \mathcal{M}$ to get the last equality.
Next, suppose that $\underset{m}{\sum} y_m^c <C$. In this case we have
\begin{eqnarray*}
C > \underset{m}{\sum} y_m^c = \underset{m}{\sum} x_m^c  = {\underset{m}{\sum} {p_m^c}}/{\mu}=\min\Big\{C,\widehat{C}^c\Big\}
\end{eqnarray*}
implying $\widehat{C}^c<C$ so that from (C3-c)  in (\ref{lambda_equn}) we obtain  $\lambda^c=0$. Thus, the slackness condition in (\ref{slackness1_equn}) is verified.

Finally, since the problem SYSTEM  in (\ref{system_equn}) is convex, the sufficiency of the KKT conditions \cite{boyd_vandenberghe04convex_optimization,bertsekas_etal03convex_optimization} imply that $(\textbf{x}^c,\textbf{y}^c)$ is (primal) optimal.
\QEDB

\section*{Multiple Link Case}
\section{System Model and Problem Definition}
\label{multiple_link_section}
In this section we extend our results to the more general setting with multiple links. We assume an example scenario with parallel links so that the users have the flexibility to off-load different amounts of rates on different links. Simultaneously, the respective link-managers have to be competitive in terms of their bids in order to maximize their respective pay-offs\footnote{Extension to a general network as in Kelly \cite{kelly97charging-rate-control} is straightforward and does not bring out any new phenomenon.}. Although it is natural to expect active participation from both users and link-managers, in the upcoming Theorem~\ref{ML_NE_theorem} we show the contrary. We will see that, when the users and the link-managers are strategic, the market collapses due to zero participation from both types of agents. This outcome is similar to the single-link case. This also establishes that the break-down in the single-link case is not due to the monopolistic nature of the supplier in the single-link setting. In Theorem~\ref{ML_NE_theorem}, alternative routes exist, and yet, the undesirable equilibrium ensues.

We begin by generalizing our notation from Section~\ref{system_model_section}.  As before we assume that there are $M$ users in the system. However,  we now generalize our earlier model by introducing $L$ parallel links. The capacity of link $\ell=1,2,\cdots,L$ is given by $C_\ell>0$. Let $x_{m\ell}$ denote the rate requested by user $m$ on link $\ell$, and let $y_{m\ell}$ be the rate the link-manager $\ell$ is willing to allocate to user $m$. Thus, $\textbf{x}_m=(x_{m1},x_{m2},\cdots,x_{mL})$ is the \emph{rate-request vector} of user $m$, and $\textbf{y}_{\ell}=(y_{1\ell},y_{2\ell},\cdots, y_{M\ell})$ is the \emph{rate-allocation vector} of link $\ell$. Let $\textbf{X}=(\textbf{x}_{m}:m=1,2,\cdots,M)$ and $\textbf{Y}=(\textbf{y}_\ell:\ell=1,2,\cdots,L)$ denote the rate-request matrix and rate-allocation matrix, respectively. The user pay-off and the link-cost functions are given by $U_m$ and $V_\ell$. As before, we assume that $U_m$ and $V_\ell$ are concave and strictly convex, respectively. In addition,  both  $U_m$ and $V_\ell$ are strictly increasing and continuously differentiable with $U'_m(0)$ finite.

The analog of the problem SYSTEM in (\ref{system_equn}) is given by  (in the sequel, the acronym ML stands for Multi-Link):

ML-SYSTEM
\begin{subequations}
\label{ML_system_equn}
\begin{align}
\label{ML_system_objective_equn}
\mbox{Maximize:}\hspace{2mm} & \underset{m}{\sum} U_m\left(\sum_\ell x_{m\ell}\right) - \sum_\ell V_\ell\left(\sum_m y_{m\ell}\right) \\
\label{ML_system_capacity_constraint_equn}
\mbox{Subject to:}\hspace{2mm}  & \sum_m y_{m\ell} \le C_\ell\ \forall \ell \\
\label{ML_system_matching_constraint_equn}
& x_{m\ell}\le y_{m\ell}, x_{m\ell}\ge0, y_{m\ell}\ge0\ \forall m,\ell.
\end{align}
\end{subequations}

\noindent
Similarly, denoting the users\textquotesingle\ and the link-managers\textquotesingle\ bid-vectors as
\begin{eqnarray*}
\textbf{p}_m &=&(p_{m1},p_{m2},\cdots,p_{mL})  \\
\bm{\beta}_\ell&=&(\beta_{1\ell},\beta_{2\ell},\cdots,\beta_{M\ell}),
\end{eqnarray*}
respectively, the analog of problem NETWORK in (\ref{network_equn}) is:

ML-NETWORK
\begin{subequations}
\label{ML_network_equn}
\begin{align}
\label{ML_multilink_net1_objective_equn}
\mbox{Maximize:}\hspace{2mm} &
\vc{\sum}{}{m,\ell}\left(
p_{m\ell} \log(x_{m\ell}) - \frac{y_{m\ell}^2}{2\beta_{m\ell}} \right) \\
\label{ML_multilink_net1_capacity_constraint_equn}
\mbox{Subject to:}\hspace{2mm}  & \sum_m y_{m\ell} \le C_\ell\ \forall \ell \\
\label{ML_multilink_net1_matching_constraint_equn}
& x_{m\ell}\le y_{m\ell}, x_{m\ell}\ge0, y_{m\ell}\ge0\ \forall m,\ell.
\end{align}
\end{subequations}
We introduce some more notation. Let $\textbf{P}=(\textbf{p}_m:m=1,2,\cdots,M)$ denote the users\textquotesingle\  bid matrix. Similarly, the link-managers\textquotesingle\  bid matrix is denoted by $\textbf{B}=(\bm{\beta}_\ell:\ell=1,2,\cdots,L)$. The network-manager sets prices $\bm{\lambda}=(\lambda_1,\lambda_2,\cdots,\lambda_L)$ and $\textbf{M}=(\bm{\mu}_m:m=1,2,\cdots,M)$ where $\bm{\mu}_m=(\mu_{m1},\mu_{m2},\cdots,\mu_{mL})$. The prices $\bm{\lambda}$ and $\textbf{M}$ are essentially the Lagrange multipliers associated with the constraints (\ref{ML_multilink_net1_capacity_constraint_equn}) and (\ref{ML_multilink_net1_matching_constraint_equn}), respectively.

We investigate the price-taking and the price-anticipating scenarios separately, as was done in the single-link setting.

\section{Price-Taking Scenario}
The mechanism under the price-taking scenario is exactly as in Section~\ref{price_taking_section} (see PTM in Section~\ref{price_taking_section}), except that now there are multiple link-managers who submit their respective bids $\bfb_\ell$ ($\ell=1,2,\cdots,L$) simultaneously.
In this setting, given the prices $(\bm{\lambda},\textbf{M})$ set by the network-manager, the pay-off to user $m$ can be written as
\begin{eqnarray}
P_m(\bm{p}_m;\bm{\mu}_m) = U_m\left(\underset{\ell}{\sum}\frac{p_{m\ell}}{\mu_{m\ell}}\right) - \underset{\ell}{\sum}p_{m\ell}.
\end{eqnarray}
Similarly, the pay-off to the link-manager $\ell$ is given by
\begin{eqnarray}
{P_{L,\ell}\Big(\bm{\beta}_\ell; (\bm{\mu}_\ell,\lambda_\ell)\Big)}
&=& - V_\ell\left(\sum_m\beta_{m\ell}\Big({\mu_{m\ell}-\lambda_\ell}\Big)\right) \nonumber \\
&& \hspace{4mm}+ \sum_m \beta_{m\ell}{\Big(\mu_{m\ell}-\lambda_\ell\Big)^2}
\end{eqnarray}
where $\bm{\mu}_\ell:=(\mu_{1,\ell},\mu_{2,\ell},\cdots,\mu_{M,\ell})$. The following are the generalizations of Definition~\ref{CE_definition} and Theorem~\ref{competitive_equilibrium_theorem}, respectively.

\begin{definition}[Competitive Equilibrium]
A vector of bids and prices $(\bm{P},\textbf{B},\bm{\lambda},\textbf{M})$ is said to constitute a competitive equilibrium if the following conditions hold:
\begin{enumerate}
\item[(\bf{C1})]
$P_m(\textbf{p}_m;\bm{\mu}_m) \ge P_m(\overline{\textbf{p}}_m;\bm{\mu}_m)\ \forall \overline{\textbf{p}}_m\ge\textbf{0}, \forall m$\\
\item[(\bf{C2})] $P_{L,\ell}\Big(\bm{\beta}_\ell;(\bm{\mu}_\ell,\lambda_\ell)\Big) \ge P_{L,\ell}\Big(\overline{\bm{\beta}}_\ell;(\bm{\mu}_\ell,\lambda_\ell)\Big) \forall \overline{\bm{\beta}}_\ell \ge\textbf{0}, \forall \ell$ \\
\item[(\bf{C3})] For each $\ell$ define $\mathcal{M}_\ell=\Big\{m:\mu_{m\ell}\ne\lambda_\ell\Big\}$ and $\widehat{C}_\ell=\sqrt{  \left(\sum_{m} p_{m\ell} \right) \left(\sum_{m\in\mathcal{M}_\ell}{\beta_{m\ell}}\right)   }$.
 Then,
 \begin{itemize}
 \item[(\bf{C3-a})]
 $\frac{p_{m\ell}}{\mu_{m\ell}} = {\beta_{m\ell}}\Big({\mu_{m\ell}-\lambda_\ell}\Big)\ \forall (m,\ell)$;

 \item[(\bf{C3-b})] $\mu_{m\ell}=\mu(\ell)$  where
\begin{eqnarray*}
\mu(\ell) = {\vc{\sum}{}{i} p_{i\ell}}\left/{\min\Big\{C_\ell,\widehat{C}_\ell\Big\}}\right.  \forall\ell, \forall m\in\mathcal{M};
 \end{eqnarray*}

\item[(\bf{C3-c})] $\lambda_\ell = \min\left\{0,\left(1-\left(\frac{C_\ell}{\widehat{C}_\ell}\right)^2\right) \frac{\vc{\sum}{}{i} p_{i\ell} }{C_\ell}\right\}\ \forall \ell.$
\end{itemize}
\QED
\end{enumerate}
\end{definition}

\begin{theorem}
\label{ML_competitive_equilibrium_theorem}
When the users and the link-managers are price-taking, there exists a competitive equilibrium. Moreover, given a competitive equilibrium $(\textbf{P},\bm{B},\bm{\lambda},\textbf{M})$, the rate matrices $\textbf{X}$ and $\textbf{Y}$, defined as $x_{m\ell}={p_{m\ell}}/{\mu_{m\ell}}$ and $y_{m\ell}=\beta_{m\ell}\Big({\mu_{m\ell}-\lambda_\ell}\Big)$ $\forall (m,\ell)$,  are optimal  for the problem ML-SYSTEM in (\ref{ML_system_equn}).
\end{theorem}
\begin{IEEEproof}
The proof is omitted since it is a straightforward extension of the proof of Theorem~\ref{competitive_equilibrium_theorem}.
\end{IEEEproof}

\section{Price-Anticipating Scenario -- Nash Formulation}
Recall that when the users and the link-managers are price-anticipating they expect that the bids submitted by them affect the prices set by the network-manager. In particular, the users and the link-managers  are aware that the prices $(\bm{\lambda}(\textbf{P},\textbf{B}),\textbf{M}(\textbf{P},\textbf{B}))$ set by the network-manager, in response to the bids (\textbf{P},\textbf{B}) submitted by the agents, are dual-optimal for the problem ML-NETWORK in (\ref{ML_network_equn}). The details of the mechanism under the Nash framework of the price-anticipating scenario is similar to PAM-N in Section~\ref{price_anticipating_section}, except that the setting now consists of multiple link-managers who submit their respective bids simultaneously.

Now, the expressions for the prices set by the network-manager is as reported in the following lemma (which is in-line with the result in Lemma~\ref{lagrange_multipliers_lemma}).

\begin{lemma}
\label{ML_lagrange_multipliers_lemma}
Given any matrix $(\textbf{P},\textbf{B})$ of users\textquotesingle\ and link-managers\textquotesingle\ bids, the prices $(\bm{\lambda}(\textbf{P},\textbf{B}),\textbf{M}(\textbf{P},\textbf{B}))$ set by the network-manager are given by, $\forall (\ell,m)$
\begin{eqnarray*}
\label{lbd_equn-2}
\lambda_\ell(\textbf{P},\textbf{B})
&=&
\left\{
\begin{array}{ll}
0 & \mbox{ if } \vc{\sum}{}{i}\sqrt{{p_{i\ell}}{\beta_{i\ell}}}\le C_\ell \\
f^{-1}_{\ell}(C_\ell) & \mbox{ otherwise, }
\end{array}
\right.
\end{eqnarray*}
where $f^{-1}_{\ell}$ denotes the inverse of the function $f_{\ell}$ which is defined as
\begin{eqnarray*}
\label{f_func_equn-2}
f_{\ell}(t)
&=& \vc{\sum}{}{i} \left(\frac{2p_{i\ell}}{t+\sqrt{t^2+4\frac{ p_{i\ell}}{\beta_{i\ell}}}}\right),
\end{eqnarray*}
and
\begin{eqnarray*}
\label{mu_equn-2}
\mu_{m\ell}(\textbf{P},\textbf{A})
&=& \frac{\lambda_\ell(\textbf{P},\textbf{A}) + \sqrt{\lambda_\ell(\textbf{P},\textbf{A})^2+4\frac{p_{m\ell}}{\beta_{m\ell}}}}{2}.
\end{eqnarray*}
\end{lemma}
\begin{IEEEproof}
Omitted since it is a straightforward extension of the proof of Lemma~\ref{lagrange_multipliers_lemma}.
\end{IEEEproof}

The pay-off to the users in the price-anticipating scenario can thus be expressed as
\begin{eqnarray}
\label{ML_user_payoff_equn}
Q_m(\textbf{p}_m,\textbf{p}_{-m},\textbf{B})
=U_m\left(\underset{\ell}{\sum}\frac{p_{m\ell}}{\mu_{m\ell}(\textbf{P},\textbf{B})}\right) - \underset{\ell}{\sum}p_{m\ell}.
\end{eqnarray}
It is possible to simplify the above expression by substituting for $\mu_{m\ell}(\textbf{P},\textbf{B})$. We then obtain an expression that is a generalization of the pay-off function in  (\ref{user_payoff_equn}) (note that $Q_m(\cdot)$ in (\ref{user_payoff_equn}) is for $L=1$). However, we now need to consider $2^L$ sub-cases depending on whether $\sum_{i}\sqrt{{p_{i\ell}}{\beta_{i\ell}}}\le C_\ell$ or otherwise, for each $\ell=1,2,\cdots,L$. For instance, for $L=2$ there are $4$ possible sub-cases. Suppose $\sum_{i}\sqrt{{p_{i1}}{\beta_{i1}}}\le C_1$ and $\sum_{i}{\sqrt{p_{i2}}{\beta_{i2}}} > C_2$. Then, denoting $\lambda_2:=\lambda_2(\textbf{P},\textbf{A})$, we have
\begin{eqnarray*}
\lefteqn{Q_m(\textbf{p}_m,\textbf{p}_{-m},\textbf{B})} \\
&=& U_m\left(\sqrt{{p_{m1}}{\beta_{m1}}}+\frac{2p_{m2}}{\lambda_2 + \sqrt{\lambda_2^2+4\frac{p_{m2}}{\beta_{m2}}}}\right) - \vc{\sum}{2}{\ell=1}p_{m\ell}.
\end{eqnarray*}

The expression for the links\textquotesingle\ pay-off function, however, comprises only two sub-cases as in (\ref{link_payoff_equn}):
\begin{eqnarray}
\lefteqn{Q_{L,\ell}(\bm{\beta}_\ell,\bm{\beta}_{-\ell},\textbf{P})} \nonumber \\
&=& \hspace*{-.1in}
\nonumber
\left\{
\begin{array}{ll}
-V_\ell\left(\vc{\sum}{}{m}\sqrt{{p_{m\ell}}{\beta_{m\ell}}}\right) + \underset{m}{\sum} p_{m\ell} \hspace{2mm} \mbox{ if } \vc{\sum}{}{i}\sqrt{{p_{i\ell}}{\beta_{i\ell}}}\le C_\ell \\
-V_\ell(C_\ell) +  \underset{m}{\sum} \frac{1}{\beta_{m\ell}} \left(\frac{2p_{m\ell}}{\lambda_\ell+\sqrt{\lambda_\ell^2+4\frac{p_{m\ell}}{\beta_{m\ell}}}}\right)^2  \mbox{ otherwise}
\end{array}
\right.\\
&&
\end{eqnarray}
where $\lambda_\ell:=\lambda_\ell(\textbf{A},\textbf{P})$. Using the above pay-off functions, the definition of Nash equilibrium in Definition~\ref{NE_definition} can be analogously extended to bid vectors $(\textbf{P},\textbf{B})$  in the multiple-link setting.

Finally, in the following theorem we report the existence and uniqueness of the Nash equilibrium $(\textbf{P}^o,\textbf{B}^o)$, which is inefficient in the sense that the market is non-functional at $(\textbf{P}^o,\textbf{B}^o)$  with users  making zero payments (i.e., $p_{m\ell}=0$) and the link-managers providing zero rate (i.e., ${\beta_{m\ell}}=0$).

\begin{theorem}
\label{ML_NE_theorem}
When the users and the link-managers are price-anticipating, the only Nash equilibrium is $(\textbf{P}^o,\textbf{B}^o)$ where $p_{m\ell}^o=0$ and $\beta_{m\ell}^o=0$\ $\forall m,\ell$.
\end{theorem}
\begin{IEEEproof}
The proof of the first part, that $(\textbf{P}^o,\textbf{A}^o)$ is an ML Nash equilibrium, is identical to the proof of the corresponding part in Theorem~\ref{pa_NE_theorem}. The uniqueness part however requires some modifications since the user pay-offs are now functions of its payments-to and signals-from all the links.

As in the proof of Theorem~\ref{pa_NE_theorem}, we begin by assuming that $(\textbf{P}^*,\textbf{B}^*)$ is an ML Nash equilibrium. Suppose that $p_{m\ell}^*>0$ for some $(m,\ell)$. For simplicity define
\begin{eqnarray*}
\delta_{m\ell}
&=&
\left\{
\begin{array}{ll}
\sqrt{{p_{m\ell}^*}{\beta_{m\ell}^*}} & \mbox{ if } \vc{\sum}{}{i}\sqrt{{p_{i\ell}^*}{\beta_{i\ell}^*}}\le C_\ell \\
\frac{2p_{m\ell}^*}{\lambda_\ell^*+\sqrt{{\lambda_\ell^*}^2+4\frac{p_{m\ell}^*}{\beta_{m\ell}^*}}} & \mbox{ otherwise}
\end{array}
\right.
\end{eqnarray*}
where $\lambda_\ell^*=\lambda_\ell(\textbf{P}^*,\textbf{B}^*)$. If $\beta_{m\ell}^*=0$, then $\delta_{m\ell}=0$ and we can write
\begin{eqnarray*}
Q_m(\textbf{p}_m^*,\mathbf{p_{-m}^*},\textbf{B}^*)
&=&U_m\left(\underset{j}{\sum}\delta_{mj}\right) - \underset{j}{\sum}p_{mj}^* \\
&=&U_m\left(\underset{j\ne\ell}{\sum}\delta_{mj}\right) - \underset{j}{\sum}p_{mj}^* \\
&<&U_m\left(\underset{j\ne\ell}{\sum}\delta_{mj}\right) - \underset{j\ne\ell}{\sum}p_{mj}^* \\
&=& Q_m(\overline{\textbf{p}}_m,\mathbf{p_{-m}^*},\textbf{B}^*)
\end{eqnarray*}
where $\overline{\textbf{p}}_m$ is such that $\overline{p}_{m\ell}=0$ while
$\overline{p}_{mj}=p_{mj}^*\ \forall j\ne\ell$. The above is a contradiction to our assumption that  $(\textbf{P}^*,\textbf{B}^*)$ is an ML Nash equilibrium. On the other hand, if $\beta_{m\ell}^*>0$, we have
\begin{eqnarray*}
Q_{L,\ell}(\bm{\beta}_\ell^*,\bm{\beta_{-\ell}^*},\textbf{P}^*)
&<& -V_\ell(0)+\underset{m}{\sum}p_{m\ell}^* \\
&=& Q_{L,\ell}(\bm{\beta}_\ell^o,\bm{\beta_{-\ell}^*},\textbf{P}^*)
\end{eqnarray*}
a contradiction. Hence, $p_{m\ell}^*=0\ \forall (m,\ell)$, i.e., $\textbf{P}^*=\textbf{0}$.

Now, suppose $\beta_{m\ell}^*>0$  for some $(m,\ell)$. Then, as in the proof of Theorem~\ref{pa_NE_theorem}, we have
\begin{eqnarray*}
Q_m(\textbf{p}_m^*,\mathbf{p_{-m}^*},\textbf{B}^*)
&=& U_m(0)\  (\mbox{since } \textbf{p}_{m}^*=\textbf{0}) \\
&<& U_m\left(\sqrt{{\overline{p}_{m\ell}}{\beta_{m\ell}^*}}\right) - \overline{p}_{m\ell}  \\
&=&Q_m(\overline{\textbf{p}}_m,\mathbf{p_{-m}^*},\textbf{B}^*)
\end{eqnarray*}
where $\overline{\textbf{p}}_m$ is such that $\overline{{p}}_{mj}=p_{mj}^*\ \forall \ell\ne j$ and $\overline{p}_{m\ell}$ satisfies
\begin{eqnarray*}
0<\overline{p}_{m\ell}\le \min\Big\{{C_\ell^2}/{\beta_{m\ell}^*},q_{m}\Big\}
\end{eqnarray*}
where $q_m$ is the maximizer of the function $h(\cdot)$ in (\ref{h_func_equn}). Thus, $\beta_{m\ell}^*=0\ \forall (m,\ell)$ (i.e., $\textbf{B}^*=\textbf{0}$) for a Nash equilibrium. This completes the proof.
\end{IEEEproof}

\section{Price-Anticipating Scenario -- Stackelberg Formulation}
\label{ML_price_anticipation}

We begin as in Section~\ref{price_anticipation_delay_section}. The network-manager announces the allocation procedure and the payment determination function. The link-suppliers then choose their bid-vectors $\bm{\beta}_\ell$ ($\ell=1,2,\cdots,L$). Following this the users  choose their bid-vectors $\bfp_m^{\bfB}$ ($m=1,2,\cdots,M$) where $\bfB = (\bm{\beta}_{\ell}: \ell = 1,2,\cdots,L)$ denotes the link-bid matrix. Let ${\bfP}^{\bfB}$ denote the user bid-matrix. The mechanism under this scenario is exactly as detailed under PAM-S in Section~\ref{price_anticipation_delay_section}, except that now there are multiple link-managers who announce their respective bids $\bfb_\ell$ simultaneously in Step-2. We then have the following generalization of the Stackelberg equilibrium.

\begin{definition}[Stackelberg Equilibrium]
\label{defn:se-ML}
A bid matrix $(\bfB,{\bfP}^{\bfB})$ is said to constitute a Stackelberg equilibrium for the game $\mathcal{G}(\{U_m\},\{V_\ell\})$ if, for all $m=1,2,\cdots, M$ and $\ell=1,2,\cdots,L$, we have
\begin{eqnarray*}
Q_m({\bfp}_m^{\bm{B}},\mathbf{p_{-m}^{\bm{B}}},\bfB) &\ge& Q_m(\overline{\bfp}_m,\mathbf{p}_{-m}^{\bm{B}},\bfB)\hspace{3mm} \forall \overline{\bfp}_m\ge0 \\
Q_{L,\ell}({\bfb}_{\ell},{\bfb}_{-\ell},{\bfP}^{\bfB})&\ge& Q_{L,\ell}(\overline{\bfb}_{\ell},{\bfb}_{-\ell},{\bfP}^{\overline{\bfB}})\hspace{3mm} \forall \overline{\bm{\beta}}\ge\textbf{0}
\end{eqnarray*}
where $\overline{\bfB}=[\overline{\bfb}_\ell\ {\bfb}_{-\ell}]$ is the bid matrix that results when link $\ell$ unilaterally deviates from ${\bfb}_\ell$ to $\overline{\bfb}_\ell$.
\QED
\end{definition}

Again, as in Section~\ref{price_anticipation_delay_section}, for simplicity we relax the capacity constraint by assuming that $C_\ell=\infty$ for all $\ell=1,2,\cdots,L$. We then have, for all $m$ and $\ell$
\begin{eqnarray}
\label{ML_user_payoff_equn}
Q_m({\bfp}_m,{\bfp}_{-m},\bfB) =U_m\left(\sum_\ell\sqrt{p_{m\ell}\beta_{m\ell}}\right) - \sum_\ell p_{m\ell} \\
\label{ML_link_payoff_equn}
Q_{L,\ell}({\bfb}_\ell,{\bfb}_{-\ell},{\bfP}) = -V_\ell\left(\sum_{m}\sqrt{p_{m\ell}\beta_{m\ell}}\right) +\sum_m p_{m\ell}.
\end{eqnarray}

From (\ref{ML_user_payoff_equn}) we see that the pay-off of user-$m$ is completely decoupled from ${\bfp}_{-m}$, the pay-off of other users. As a result, given $\bfB$, the equilibrium strategy of user-$m$ can be simply expressed as
\begin{eqnarray}
\label{ML_pB_equn}
{\bfp}_m^{\bfB} = \argmax_{\bfp_m} \left(U_m\left(\sum_\ell\sqrt{p_{m\ell}\beta_{m\ell}}\right) - \sum_\ell p_{m\ell}\right).
\end{eqnarray}
The following lemma is then a generalization Lemma~\ref{rbeta_lemma}.

\begin{lemma}
\label{ML_pml_lemma}
For a given $\bfB$ we have, for all $(m,\ell)$
\begin{eqnarray}
\label{ML_pml_equn}
p_{m\ell}^{\bfB}
&=&
\left\{
\begin{array}{ll}
\frac{\beta_{m\ell} r_m(\bfB)^2}{\left(\sum_k \beta_{m k}\right)^2} & \mbox{ if } \sum_k \beta_{m k}>0  \\\\
0 & \mbox{ otherwise,}
\end{array}
\right.
\end{eqnarray}
where $r_m(\bfB)=r_m({\bfb}_{\ell},{\bfb}_{-\ell})$ is the solution to the equation $U_m'(r)=2r/\left(\sum_k \beta_{m k}\right)$. 
\end{lemma}
\begin{IEEEproof}
The optimality equation for the problem in (\ref{ML_pB_equn}) is given by
\begin{eqnarray*}
U_m'\left(\sum_k \sqrt{p_{mk}\beta_{mk}}\right) \frac{\sqrt{\beta_{m\ell}}}{2\sqrt{p_{m\ell}}}-1 = 0 && \mbox{ if } \beta_{m\ell}>0 \\
p_{m\ell} = 0 && \mbox{ if } \beta_{m\ell}=0.
\end{eqnarray*}
It is easy to check that $p_{m\ell}^{\bfB}$ in (\ref{ML_pml_equn}) satisfies the above conditions, thus verifying  first part of the lemma. For the second part,  note that $\sum_k \beta_{m k}=0$ implies $\beta_{m\ell}=0$ for all $\ell$ (since $\beta_{m\ell}$ are non-negative). Thus, the problem in  (\ref{ML_pB_equn}) reduces to ${\bfp}_m^{\bfB} = \argmax_{\bfp_m} \left(- \sum_\ell p_{m\ell}\right)$, the (non-negative) solution to which is simply given by $p_{m\ell}^{\bfB}=0$ for all $\ell$.
\end{IEEEproof}

Substituting for ${\bfP}^{\bfB}$ in (\ref{ML_link_payoff_equn}), the link pay-off functions can be expressed as
\begin{eqnarray}
\label{ML_Sl_equn}
\lefteqn{S_\ell({\bfb}_{\ell},{\bfb}_{-\ell}):=Q_{L,\ell}({\bfb}_\ell,{\bfb}_{-\ell},{\bfP}^{\bfB})} \nonumber\\
&&=  -V_\ell\left(\sum_{m}\sqrt{p_{m\ell}^{\bfB}\beta_{m\ell}}\right) +\sum_m p_{m\ell}^{\bfB} \nonumber\\
&&= -V_{\ell}\left(\sum_m \frac{\beta_{m\ell}r_m({\bfb}_{\ell},{\bfb}_{-\ell})}{\sum_k \beta_{mk}}\right) + \sum_m \frac{\beta_{m\ell}r_m({\bfb}_{\ell},{\bfb}_{-\ell})^2}
{\left(\sum_k\beta_{mk}\right)^2} \nonumber\\
\end{eqnarray}
for $\ell=1,2,\cdots,L$.
We denote the  game played by the links alone with the above pay-off functions as $\mathcal{G}_L(\{U_m\},\{V_\ell\})$, while the Stackelberg game played by all agents is denoted $\mathcal{G}(\{U_m\},\{V_\ell\})$.
\begin{definition}
$\bfB^*=[\bfb_\ell^* \bfb_{-\ell}^*]$ is said to constitute a Nash equilibrium for the game $\mathcal{G}_L(\{U_m\},\{V_\ell\})$ if
\begin{eqnarray*}
S_{\ell}({\bfb}_\ell^*,{\bfb}_{-\ell}^*) &\ge&
S_{\ell}({\bfb}_\ell,{\bfb}_{-\ell}^*).
\end{eqnarray*}
for all ${\bfb}_{\ell}\ge \bf{0}$ and $\ell=1,2,\cdots,L$.
\QED
\end{definition}
The following lemma is straightforward.

\begin{lemma}
\label{lem:psne-implies-se}
If $\bfB^*$ is a Nash equilibrium for the game $\mathcal{G}_L(\{U_m\},\{V_\ell\})$ then $(\bfB^*,\bfP^{\bfB^*})$ is a Stackelberg equilibrium for the game $\mathcal{G}(\{U_m\},\{V_\ell\})$.
\end{lemma}

Unlike the single-link case (recall Lemma~\ref{stackelberg-existence-lemma}), existence of a Stackelberg equilibrium for the game $\mathcal{G}(\{U_m\},\{V_\ell\})$ under general user pay-offs is still open. This is because existence of a (pure strategy) Nash equilibrium for the game $\mathcal{G}_L(\{U_m\},\{V_\ell\})$ is still open\footnote{Tian \cite[Th.~3.1]{tian2015existence} provides a necessary and sufficient condition for existence of pure strategy Nash equilibria in games with compact action spaces. We have not been able to verify whether Tian's condition holds for the game $\mathcal{G}_L(\{U_m\},\{V_\ell\})$.}.

In the special case of linear user pay-offs, $U_m(x_m) = c_m x_m$, the pay-off functions in (\ref{ML_Sl_equn}) for the game $\mathcal{G}_L(\{U_m\},\{V_\ell\})$ get decoupled from the actions of other players. The analysis is greatly simplified and one can assert that a (pure strategy) Nash equilibrium exists for the game $\mathcal{G}_L(\{U_m\},\{V_\ell\})$, and by Lemma \ref{lem:psne-implies-se}, a corresponding Stackelberg equilibrium in the sense of Definition \ref{defn:se-ML} exists. 

\subsection{Stackelberg Equilibrium for Linear User Pay-offs}
Consider linear pay-off functions $\{U_m\}$, where ${U}_m(x_m)=c_mx_m$ with $a_m>0$. Now, given a bid matrix $\bfB=[\bfb_\ell,~ \bfb_{-\ell}]$, using (\ref{ML_pml_equn}) we can write
\begin{eqnarray}
\label{ML_rm_equn}
{r}_m(\bfB) = \frac{\left(\sum_k \beta_{m k}\right){U}_m'({r}_m(\bfB))}{2} = \frac{\left(\sum_k \beta_{m k}\right)c_m}{2}.
\end{eqnarray}
Substituting the above in (\ref{ML_Sl_equn}) simplifies the link pay-off functions:
\begin{eqnarray*}
{S}_\ell({\bfb}_{\ell},{\bfb}_{-\ell})
= -V_{\ell}\left(\sum_m \frac{\beta_{m\ell}c_m}{2}\right) + \sum_m \frac{\beta_{m\ell}c_m^2}{4}
\end{eqnarray*}
Thus, when the pay-offs are linear, the pay-off of link-manager $\ell$ depends solely on the action $\bfb_\ell=(\beta_{m\ell}:m=1,\cdots,M)$ chosen by him. Thus, for $\bfB^*=[\bfb_\ell^*,~ {\bfb}_{-\ell}^*]$ to be a Nash equilibrium the following should hold:
\begin{eqnarray}
\label{ML_beta_star_set_equn}
{\bfb}_\ell^*\in \argmax_{\bfb_{\ell} \geq \bm{0}} \left\{{-V_{\ell}\left(\sum_m \frac{\beta_{m\ell}c_m}{2}\right) + \sum_m \frac{\beta_{m\ell}c_m^2}{4}}\right\}.
\end{eqnarray}
for all $\ell=1,2,\cdots,L$.

Without loss of generality, assume that $c_1=\max_m\{c_m\}$. Then, the solution to the above problem is given by
\begin{eqnarray}
\label{ML_betaml_star_equn}
\beta_{m\ell}^* =
\left\{
\begin{array}{ll}
\frac{2}{c_m} v_\ell^{-1}\left(\frac{c_m}{2}\right) & \mbox{ if } m=1 \\
0 & \mbox{ otherwise,}
\end{array}
\right.
\end{eqnarray}
for all $(m,\ell)$, where $v_\ell(x):=V_\ell'(x)$. Substituting for $\beta_{m\ell}^*$  in (\ref{ML_pml_equn}) yields
\begin{eqnarray}
\label{ML_pml_star_equn}
p_{m\ell}^{\bfB^*} =
\left\{
\begin{array}{ll}
\frac{c_m}{2} v_\ell^{-1}\left(\frac{c_m}{2}\right) & \mbox{ if } m=1 \\
0 & \mbox{ otherwise. }
\end{array}
\right.
\end{eqnarray}
Thus, the  rate allocated to user $m$ on link-$\ell$ at equilibrium is given by
\begin{eqnarray}
\label{ML_xml_equn}
x_{m\ell}^{\bfB^*} &=& \sqrt{ p_{m\ell}^{\bfB^*} \beta_{m\ell}^{\bfB^*}}  \nonumber \\
&=&
\left\{
\begin{array}{ll}
v_\ell^{-1}\left(\frac{c_m}{2}\right) & \mbox{ if } m=1 \\
0 & \mbox{ otherwise. }
\end{array}
\right.
\end{eqnarray}
Further, recalling (\ref{ML_rm_equn}), the total rate allocated to user $m$ at equilibrium can be written as
\begin{eqnarray}
\label{ML_xrate_equn}
\sum_{\ell}x_{m\ell}^{{\bfB}^*}
&=&  \sum_\ell \sqrt{ p_{m\ell}^{\bfB^*} \beta_{m\ell}^{\bfB^*}} = {r}_m({\bfB}^*)\nonumber \\
&=&
\left\{
\begin{array}{ll}
\sum_\ell v_\ell^{-1}\left(\frac{c_m}{2}\right)  & \mbox{ if } m=1 \\\\
0  & \mbox{ otherwise. }
\end{array}
\right.
\end{eqnarray}
Similarly, the total rate served by the link-manager $\ell$ at equilibrium is given by
\begin{eqnarray}
\label{ML_yrate_equn}
\sum_{m}y_{m\ell}^{{\bfB}^*}
=  \sum_m \sqrt{ p_{m\ell}^{\bfB^*} \beta_{m\ell}^{\bfB^*}}
= v_{\ell}^{-1}\left(\frac{c_1}{2}\right)
\end{eqnarray}
The foregoing establishes that a Stackelberg equilibrium exists for the game $\mathcal{G}(\{U_m\},\{V_\ell\})$ where the $U_m$ are linear pay-offs.

\subsection{Bound on Efficiency for Linear User Pay-offs}
As in the single link case, we define the efficiency of  a Stackelberg equilibrium $(\bfB^*,\bfP^{\bfB^*})$ as
\begin{eqnarray}
\label{ML_PoA_equn}
\lefteqn{\mathcal{E}(\{U_m\};\{V_\ell\}) =} \hspace{8mm}\nonumber\\
&& \frac{\sum_m U_m\left(\sum_{\ell}x_{m\ell}^{\bfB^*}\right) - \sum_{\ell}V_\ell\left(\sum_m x_{m\ell}^{\bfB^*}\right)}{\sum_m U_m\left(\sum_{\ell}x_{m\ell}^{s}\right) - V\left(\sum_m x_{m\ell}^{s}\right)}
\end{eqnarray}
where $\{x_{m\ell}^{\bfB^*}\}$ denotes the rates allocated at  Stackelberg equilibrium while $x_{m\ell}^s$ are the social optimum rates (obtained by solving ML-SYSTEM in (\ref{ML_system_equn})).

The following result is a generalization of Theorem~\ref{poa_theorem}.

\begin{theorem}
\label{ML_poa_theorem}
Fix a collection of link-cost functions $\{V_\ell(\cdot)\}$. For any set of linear user pay-offs $\{U_m\}$, we have
\begin{eqnarray}
\label{ML_poa_bound_equn}
\mathcal{E}(\{U_m\};\{V_\ell\}) \ge \inf_{c>0}\ \frac{ \sum_\ell \Big(cv_\ell^{-1}(\frac{c}{2})-V_\ell(v_\ell^{-1}(\frac{c}{2}))\Big)}{\sum_\ell\Big( c v_\ell^{-1}(c)-V_\ell(v_\ell^{-1}(c))\Big)}
\end{eqnarray}
where $v_\ell(\cdot):=V_\ell'(\cdot)$ for  $\ell=1,2,\cdots,L$.
\end{theorem}
\begin{IEEEproof}
The proof is along the lines of the proof of Theorem~\ref{poa_theorem}. For completeness we however present the details again here. Consider linear pay-offs of the form $U_m(x_m)=c_mx_m$ where $c_m>0$ ($m=1,2,\cdots,M$). Without loss of generality, assume that $c_1=\max_m \{c_m\}$. Then, recalling (\ref{ML_xml_equn}) and (\ref{ML_yrate_equn}), the utility at Stackelberg equilibrium can be written as
\begin{eqnarray}
\label{ML_stackelberg_utility}
\lefteqn{\mbox{Stackelberg utility} } \hspace{10mm}\nonumber\\
&=& \sum_m U_m\left(\sum_{\ell}x_{m\ell}^{\bfB^*}\right) - \sum_{\ell}V_\ell\left(\sum_m x_{m\ell}^{\bfB^*}\right) \nonumber \\
&=&  U_1\left(\sum_{\ell}v_\ell^{-1}\left(\frac{c_1}{2}\right)\right) - \sum_{\ell}V_\ell\left(v_\ell^{-1}\left(\frac{c_1}{2}\right)\right) \nonumber \\
&=& c_1 \sum_{\ell}v_\ell^{-1}\left(\frac{c_1}{2}\right) - \sum_{\ell}V_\ell\left(v_\ell^{-1}\left(\frac{c_1}{2}\right)\right) \nonumber \\
&=& \sum_\ell \Big(c_1 v_\ell^{-1}\left(\frac{c_1}{2}\right)-V_\ell\left(v_\ell^{-1}\left(\frac{c_1}{2}\right)\right)\Big).
\end{eqnarray}

Next, the social optimal utility is obtained by solving
\begin{eqnarray*}
\max_{\bf{X} \geq \bf{0}} \left\{ \underset{m}{\sum} U_m\left(\sum_\ell x_{m\ell}\right) - \sum_\ell V_\ell\left(\sum_m y_{m\ell}\right)  \right\}.
\end{eqnarray*}
Substituting for the linear pay-off functions and rearranging, the above problem can be alternatively expressed as
\begin{eqnarray*}
\max_{\bf{X} \geq \bf{0}} \left\{ \sum_\ell \left(\sum_m c_m x_{m\ell} - V_\ell\left(\sum_m x_{m\ell}\right) \right) \right\}.
\end{eqnarray*}
Thus, for each $\ell$ the optimal rate allocation $x_{m\ell}^S$ is given by
\begin{eqnarray*}
x_{m\ell}^{s} &=&
\left\{
\begin{array}{ll}
v_\ell^{-1}(a_1) & \mbox{ if } m=1 \\
0 & \mbox{ otherwise.}
\end{array}
\right.
\end{eqnarray*}
Thus, the social optimal utility is given by
\begin{eqnarray}
\label{ML_social_utility}
\mbox{Social utility} = \sum_\ell\Big( c_1 v_\ell^{-1}(c_1)-V_\ell\left(v_\ell^{-1}(c_1)\right)\Big).
\end{eqnarray}

From (\ref{ML_stackelberg_utility}) and (\ref{ML_social_utility}) we have (replacing $c_1$ by $c$)
\begin{eqnarray*}
\mathcal{E}(\{U_m\};\{V_\ell\}) \ge  \frac{ \sum_\ell \Big(cv_\ell^{-1}(\frac{c}{2})-V_\ell(v_\ell^{-1}(\frac{c}{2}))\Big)}{\sum_\ell\Big( c v_\ell^{-1}(c)-V_\ell(v_\ell^{-1}(c))\Big)}.
\end{eqnarray*}
The worst case bound on efficiency can be obtained by taking infimum over all $c>0$.
\end{IEEEproof}

\vspace{-2mm}
\subsection{Efficiency Bound for Polynomial Link Costs}

Applying Theorem~\ref{ML_poa_theorem} to polynomial link-cost functions, we obtain results analogous to those in Section~\ref{polynomial_subsection}. For instance, suppose all the link-cost functions are quadratic, i.e., $V_\ell(x) = a_\ell x^2$ where $a_\ell>0$ for $\ell=1,2,\cdots,L$. In this case we have, $v_\ell(x)=2a_\ell x$ so that $v_\ell^{-1}(y)=\frac{y}{2a_{\ell}}$. Thus, the bound on efficiency can be written as
\begin{eqnarray*}
\mathcal{E}(\{U_m\};\{V_\ell\})
&\ge& \inf_{c>0} \frac{\sum_\ell\left(c\frac{c}{4a_\ell} - V_{\ell}(\frac{c}{4a_\ell})\right)}{\sum_\ell\left(c \frac{c}{2a_\ell} - V_{\ell}(\frac{c}{2a_\ell})\right)} \\
&=& \inf_{c>0} \frac{\sum_\ell \left(c\frac{c}{4a_\ell} - a_\ell(\frac{c}{4a_\ell})^2\right)}{\sum_\ell \left(c \frac{c}{2a_\ell} - a_\ell(\frac{c}{2a_\ell})^2\right)} = \frac{3}{4}.
\end{eqnarray*}
The above bound is identical to that obtained for the single link case in Section~\ref{polynomial_subsection}.

Similarly, the bound on efficiency when the link costs are cubic (i.e., $V_\ell(x)=a_\ell x^3$) is $\mathcal{E}(\{U_m\};\{V_\ell\})\ge \frac{5}{4\sqrt{2}}$. In general, for polynomial link costs of the form  $V_\ell(x)=a_\ell x^n$ ($n\ge2$) we have
\begin{eqnarray*}
\mathcal{E}(\{U_m\};\{V_\ell\}) &\ge& \left(\frac{1}{2}\right)^{\frac{n}{n-1}}\frac{2n-1}{n-1}.
\end{eqnarray*}
which converges to $1$ as $n\rightarrow\infty$.
Thus, even in the multi-link setting, the efficiency approaches $1$ as $n \rightarrow \infty$ for polynomial link-cost functions.

\vspace{-2mm}
\subsection{Worst-Case Bound on Efficiency for Linear User Pay-offs}
As in single link case, the worst-case bound on efficiency can be arbitrarily close to $0$ for pathological link-cost functions. To see this, assume that $V_\ell=V$ for all $\ell$. Then, the bound on efficiency reduces to single link bound in (\ref{poa_bound_equn}):
\begin{eqnarray*}
\mathcal{E}(\{U_m\};\{V_\ell\}) &\ge& \inf_{c>0}\ \frac{c v^{-1}(\frac{c}{2})-V(v^{-1}(\frac{c}{2}))}{c v^{-1}(c)-V(v^{-1}(c))}.
\end{eqnarray*}
Now, applying the arguments in Section~\ref{worst_case_bound}, we can identify a sequence of link-cost functions $V^{(n)}$, $n\ge1$, such that the corresponding sequence of efficiency bounds converges to zeros as $n\rightarrow\infty$.

\end{document}